\documentclass[amssymb,aps,showkeys,superscriptaddress,prb]{revtex4}
\usepackage{graphicx}
\usepackage{dcolumn}
\usepackage{color}
\usepackage{bm}

\begin{document}


\title{Insights into nature of magnetization plateaus of~a~nickel complex [Ni$_4$($\mu$-CO$_3$)$_2$(aetpy)$_8$](ClO$_4$)$_4$ from~a~spin-1 Heisenberg diamond cluster}
\author{Katar\'ina Kar\v{l}ov\'a}
\affiliation{Department of Theoretical Physics and Astrophysics, Faculty of Science, P.~J.~\v{S}af\'arik University, Park Angelinum 9, 040\,01 Ko\v{s}ice, Slovakia}
\author{Jozef Stre\v{c}ka}
\email{jozef.strecka@upjs.sk}
\affiliation{Department of Theoretical Physics and Astrophysics, Faculty of Science, P.~J.~\v{S}af\'arik University, Park Angelinum 9, 040\,01 Ko\v{s}ice, Slovakia}
\author{Jozef Hani\v{s}}
\affiliation{Department of Theoretical Physics and Astrophysics, Faculty of Science, P.~J.~\v{S}af\'arik University, Park Angelinum 9, 040\,01 Ko\v{s}ice, Slovakia}
\author{Masayuki Hagiwara}
\affiliation{Center for Advanced High Magnetic Field Science, Graduate School of Science, Osaka University, Toyonaka, Osaka 560-0043, Japan}

\date{\today}

\begin{abstract}
Magnetic and magnetocaloric properties of a spin-1 Heisenberg diamond cluster with two different coupling constants are investigated with the help of an exact diagonalization based on the Kambe's method, which employs a local conservation of composite spins formed by spin-1 entities located in opposite corners of a diamond spin cluster. It is shown that the spin-1 Heisenberg diamond cluster exhibits several intriguing quantum ground states, which are manifested in low-temperature magnetization curves as intermediate plateaus at 1/4, 1/2 and 3/4 of the saturation magnetization. Besides, the spin-1 Heisenberg diamond cluster may also exhibit an enhanced magnetocaloric effect, which may be relevant for a low-temperature refrigeration achieved through the adiabatic demagnetization. It is evidenced that the spin-1 Heisenberg diamond cluster with the antiferromagnetic coupling constants $J_1$/$k_{\rm B}$ =  41.4~K and $J_2$/$k_{\rm B}$ = 9.2~K satisfactorily reproduces a low-temperature magnetization curve recorded for the tetranuclear nickel complex [Ni$_4$($\mu$-CO$_3$)$_2$(aetpy)$_8$](ClO$_4$)$_4$ (aetpy = 2-aminoethyl-pyridine) including a size and position of intermediate plateaus detected at 1/2 and 3/4 of the saturation magnetization. A microscopic nature of fractional magnetization plateaus observed experimentally is clarified and interpreted in terms of valence-bond crystal with either a single or double valence bond. It is suggested that this frustrated magnetic molecule can provide a prospective cryogenic coolant with the maximal isothermal entropy change $-\Delta S_M = 10.6$ J.K$^{-1}$.kg$^{-1}$ in a temperature range below 2.3~K.
\end{abstract}

\maketitle

\section{Introduction}
Molecular-based magnetic materials have attracted a considerable research interest over the past few decades, because they provide perspective building blocks for a development of new generation of nanoscale devices with a broad application potential \cite{carl86,kahn93,mill02,siek17}. Small magnetic molecules composed a few exchange-coupled spin centers might for instance serve for the rational design of high-density storage devices \cite{gatt03} and various spintronic devices \cite{roch05,boga08,urda11}. Another intriguing feature of a special class of molecular magnetic materials with an extremely slow magnetic relaxation, which are commonly referred to as single-molecule magnets, is their possible implementation for developing novel platform for a quantum computation and quantum information processing \cite{leue01,teja01,timc11,troi11,esca18,gait19,guwu20}.

Appearance of plateaus in low-temperature magnetization curves of molecular magnetic materials at rational values of the magnetization represents other fascinating topical issue of current research interest, which can be experimentally easily validated due to a recent development of high-field facilities \cite{kind10,lipe12,zher13,hagi13,maed14,bear18,hahn19,mich20}. The magnetization plateaus often bear evidence of unconventional quantum states of matter theoretically predicted by the respective quantum Heisenberg spin models (see Ref. \cite{taki11} and references cited therein). It should be pointed out, however, that the underlying mechanism for formation of intermediate magnetization plateau does not necessarily need to be of a purely 'quantum' origin, but it may sometimes have 'classical' character. The 'classical' plateau is a simple adiabatic continuation of a commensurate classical spin state realized in the Ising limit that is of course being subject to a quantum reduction of the local magnetization caused by quantum fluctuations, while the purely 'quantum' plateau relates to a massive quantum spin state with an energy gap that does not have any classical counterpart \cite{taki11,hida05,cole13,karl17,karl20}. 
 
Naturally, the most comprehensively understood are nowadays rational magnetization plateaus of the simplest molecular materials, which consist of well isolated magnetic molecules involving just a few spin centers coupled through antiferromagnetic exchange interactions. High-field measurements performed at sufficiently low temperatures have for instance evidenced presence of intermediate magnetization plateau(s) for the dinuclear nickel complex \{Ni$_2$\} as an experimental realization of the spin-1 Heisenberg dimer \cite{naru98,stre05,stre08}, the dinuclear nickel-copper complex \{NiCu\} as an experimental realization of the mixed spin-(1,1/2) Heisenberg dimer \cite{hagi99}, the trinuclear copper \{Cu$_3$\} and nickel \{Ni$_3$\} complexes as experimental realizations of the spin-1/2 and spin-1 Heisenberg triangles \cite{choi08,pono15,chat19}, the oligonuclear compound \{Mo$_{12}$Ni$_4$\} as an experimental realization of the spin-1 Heisenberg tetrahedron \cite{mull00,post05,schn06,zbli08}, the pentanuclear copper complex \{Cu$_5$\} as an experimental realization of the spin-1/2 Heisenberg hourglass cluster \cite{nath13,szal20}, the hexanuclear vanadium compounds \{V$_6$\} as experimental realizations of two weakly coupled spin-1/2 Heisenberg triangles \cite{luba02,kowa20}, the hexanuclear copper compounds \{Cu$_6$\} as experimental realizations of the spin-1/2 Heisenberg edge-shared tetrahedra \cite{fuji18,stre18,furr20}, etc.

Recently, significant advances have been also achieved in the design of molecular magnets providing prospective coolants for the magnetic refrigeration technology in a low- and ultra low-temperature range, where they offer a more energy-efficient, cost-effective and environmentally friendly alternative with respect to traditional refrigeration technologies based on vapour-compression technique or $^3$He-$^4$He dilution-refrigerator method \cite{evan10,shar13,garl13,liuj14,zhen14}. The magnetic cooling takes advantage of a thermal response of magnetic materials with respect to the variation of external magnetic field, which is traditionally denoted as the magnetocaloric effect (MCE). A decrease in temperature caused by the adiabatic demagnetization of a magnetic material is referred to as the conventional MCE, while a rise of temperature during the adiabatic demagnetization is contrarily referred to as the inverse MCE. The adiabatic temperature change and the isothermal change of entropy are two most important characteristics of the magnetic coolants, which basically depend on the magnetic-field change and the initial temperature \cite{evan10,shar13,garl13,liuj14,zhen14}. Although all magnetic substances display a certain magnetocaloric response, only a few molecular-based compounds possess sufficiently large isothermal change of the entropy and the adiabatic temperature change in order to be regarded as prospective magnetic refrigerants such as the molecular nanomagnets with the spin-enhanced \cite{evan05,evan09} or frustration-enhanced \cite{affr04,shar14,fuxu14,pine16} magnetocaloric features. It should be nevertheless mentioned  that the isothermal entropy change of molecular nanomagnets composed exclusively from transition-metal ions other than Mn$^{2+}$ and/or  Fe$^{3+}$ \cite{liuj14,fitt19} just rarely exceeds the value $\Delta S_M = 10$ J. K$^{-1}$. kg$^{-1}$, which may be regarded as a benchmark for the enhanced MCE.

In the present work we will examine in particular magnetic and magnetocaloric properties of the spin-1 Heisenberg diamond spin cluster, which is inspired by a magnetic structure of butterfly-tetrameric nickel complex [Ni$_4$($\mu$-CO$_3$)$_2$(aetpy)$_8$](ClO$_4$)$_4$ (aetpy = 2-aminoethyl-pyridine) \cite{escu98} hereafter abbreviated as \{Ni$_4$\}. The original study of structural and magnetic properties has evidenced a spin-frustrated character of the tetranuclear complex \{Ni$_4$\}, which arises from a competition between two different antiferromagnetic exchange interactions between constituent spin-1 Ni$^{2+}$ magnetic ions \cite{escu98}. The posterior high-field magnetization measurements has corroborated a highly frustrated character of the tetranuclear magnetic molecule \{Ni$_4$\}, which gives rise to a peculiar low-temperature magnetization curve involving two rational magnetization plateaus located at 1/2 and 3/4 of the saturation magnetization \cite{hagi06}. An exact nature of the experimentally observed magnetization plateaus along with basic magnetocaloric characteristics of the underlying spin-1 Heisenberg diamond spin cluster will be the main subject of the present article.

The organization of this paper is as follows. The spin-1 Heisenberg diamond cluster will be introduced together with basic steps of its exact analytical treatment based on the exact diagonalization method in Sec. \ref{model}. The most interesting theoretical results for the ground-state phase diagram, magnetization process and magnetocaloric properties of the spin-1 Heisenberg diamond cluster will be comprehensively investigated in Sec. \ref{theory}. The available experimental magnetization data for the tetranuclear complex \{Ni$_4$\} will be thoroughly interpreted in Sec. \ref{experiment} within the framework of the studied model, which will be additionally used for obtaining respective theoretical implications for its magnetocaloric properties not reported experimentally hitherto. Finally, the most important findings and future outlooks will be presented in Sec. \ref{conclusion}.

\section{Spin-1 Heisenberg diamond cluster}
\label{model}

\begin{figure}
\begin{center}
\includegraphics[width=0.3\textwidth]{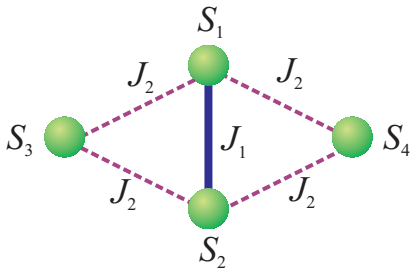}
\end{center}
\vspace{-0.5cm}
\caption{A schematic illustration of the spin-1 Heisenberg diamond cluster, which involves two different exchange interactions $J_1$ and $J_2$ along a shorter diagonal and sides of a diamond spin cluster depicted by thick solid and thin broken lines, respectively.}
\label{fig1}       
\end{figure}

Let us consider the spin-1 Heisenberg diamond cluster in a magnetic field, which is schematically illustrated in Fig. \ref{fig1} and mathematically defined through the Hamiltonian:
\begin{eqnarray}
	\mathbf{\hat{\mathcal{H}}} = J_1\mathbf{\hat{S}}_1\cdot\mathbf{\hat{S}}_2 + J_2\left(\mathbf{\hat{S}}_1 + \mathbf{\hat{S}}_2\right)\cdot\left(\mathbf{\hat{S}}_3 + \mathbf{\hat{S}}_4\right) - h \sum_{i=1}^{4} \hat{S}_i^z.
\label{hami}
\end{eqnarray}
The model Hamiltonian (\ref{hami}) aims at describing the magnetochemistry of tetranuclear nickel complex \{Ni$_4$\} with a magnetic structure of the butterfly tetramer \cite{escu98,hagi06}, which can be alternatively viewed as a diamond spin cluster involving two different exchange interactions $J_1$ and $J_2$ along its shorter diagonal and sides schematically shown in Fig.~\ref{fig1} by thick solid and thin broken lines, respectively. The last term $h = g \mu_{\rm B} B$ is the standard Zeeman's term associated with the external magnetic field $B$, which directly incorporates in its definition the Land\'e $g$-factor and the Bohr magneton $\mu_{\rm B}$.

By introducing two composite spin operators $\mathbf{\hat{S}}_{12} = \mathbf{\hat{S}}_1 + \mathbf{\hat{S}}_2$ and $\mathbf{\hat{S}}_{34} = \mathbf{\hat{S}}_3 + \mathbf{\hat{S}}_4$ within the Kambe coupling scheme \cite{kamb50,sinn70} together with the total spin operator $\mathbf{\hat{S}}_T = \mathbf{\hat{S}}_{12} + \mathbf{\hat{S}}_{34}$ and its $z$-component $\hat{S}_T^z = \sum_{i=1}^{4} \hat{S}_i^z$ one can rewrite the Hamiltonian (\ref{hami}) into the following equivalent form:
\begin{eqnarray}
\mathbf{\hat{\mathcal{H}}} = \frac{J_1}{2}\left(\mathbf{\hat{S}}_{12}^2 - 4 \right)
	 +\frac{J_2}{2}\left(\mathbf{\hat{S}}_{T}^2 - \mathbf{\hat{S}}_{12}^2 - \mathbf{\hat{S}}_{34}^2\right) - h\hat{S}_{T}^{z}.
\label{H3}
\end{eqnarray}
It can be easily verified that all spin operators $\mathbf{\hat{S}}_{12}^2$, $\mathbf{\hat{S}}_{34}^2$, $\mathbf{\hat{S}}_T^2$ and $\hat{S}_T^z$ entering on the right-hand side of Eq.~(\ref{H3}) commute with the Hamiltonian, which allows one to express energy eigenvalues in terms of the respective quantum spin numbers:
\begin{eqnarray}
	E (S_T, S_{12}, S_{34}, S_T^z)= \frac{J_1 - J_2}{2}S_{12}\left(S_{12} + 1\right) - \frac{J_2}{2}S_{34}\left(S_{34} + 1\right) 
	  + \frac{J_2}{2}S_T\left(S_T + 1\right) - 2J_1 - h S_T^z.
\label{HE}
\end{eqnarray}
A full energy spectrum can be obtained from Eq. (\ref{HE}) after considering all available combinations of the quantum spin numbers $S_{12} = 0, 1, 2$ and $S_{34} = 0,1,2$ together with the composition rules for the total spin angular momentum $S_{T} = \vert S_{12} - S_{34} \vert, \vert S_{12} - S_{34} \vert + 1, \cdots , S_{12} + S_{34}$ and its $z$-component $S_T^z = -S_T, -S_T +1, ... , S_T$ according to the Kambe coupling scheme \cite{kamb50,sinn70}. For completeness, all energy eigenvalues assigned to allowed combinations of the quantum spin numbers $S_T$, $S_{12}$, $S_{34}$ and $S_T^z$ are listed in Tab.~\ref{t1}. 

\begin{table}
	\centering
	\caption{A complete energy spectrum of the spin-1 Heisenberg diamond cluster. Each energy eigenvalue is assigned to a given set of the quantum spin numbers $S_T$, $S_{12}$, $S_{34}$, $S_T^z$.}
	\label{t1}
	\begin{tabular}{@{}|cccc|c|cccc|c|@{}}
		\hline
		$S_T$ & $S_{12}$ & $S_{34}$ & $S_T^z$ & Energy & $S_T$ & $S_{12}$ & $S_{34}$ & $S_T^z$ & Energy\\
		\hline
		$0$ & $0$ & $0$ & $0$ & $-2J_1$                  & $3$ & $2$ & $1$ & $\pm3$ & $J_1 + 2J_2 \mp 3h$ \\
		$1$ & $1$ & $0$ & $0$ & $-J_1$                   & $1$ & $1$ & $2$ & $0$ & $-J_1 - 3J_2$ \\
		$1$ & $1$ & $0$ & $\pm1$ & $-J_1 \mp h$          & $1$ & $1$ & $2$ & $\pm1$ & $-J_1 - 3J_2 \mp h$ \\
		$1$ & $0$ & $1$ & $0$ & $-2J_1$                  & $2$ & $1$ & $2$ & $0$ & $-J_1 - J_2$ \\
		$1$ & $0$ & $1$ & $\pm1$ & $-2J_1 \mp h$      	 & $2$ & $1$ & $2$ & $\pm1$ & $-J_1 - J_2 \mp h$ \\
		$0$ & $1$ & $1$ & $0$ & $-J_1 -2J_2$          	 & $2$ & $1$ & $2$ & $\pm2$ & $-J_1 - J_2 \mp 2h$ \\
		$1$ & $1$ & $1$ & $0$ & $-J_1 -J_2$            	 & $3$ & $1$ & $2$ & $0$ & $-J_1 + 2J_2$ \\
		$1$ & $1$ & $1$ & $\pm1$ & $-J_1 -J_2 \mp h$  	 & $3$ & $1$ & $2$ & $\pm1$ & $-J_1 + 2J_2 \mp h$ \\
		$2$ & $1$ & $1$ & $0$ & $-J_1 +J_2$           	 & $3$ & $1$ & $2$ & $\pm2$ & $-J_1 + 2J_2 \mp 2h$ \\
		$2$ & $1$ & $1$ & $\pm1$ & $-J_1 +J_2 \mp h$  	 & $3$ & $1$ & $2$ & $\pm3$ & $-J_1 + 2J_2 \mp 3h$ \\
		$2$ & $1$ & $1$ & $\pm2$ & $-J_1 +J_2 \mp 2h$ 	 & $0$ & $2$ & $2$ & $0$ & $J_1 - 6J_2$ \\
		$2$ & $2$ & $0$ & $0$ & $J_1$ 									 & $1$ & $2$ & $2$ & $0$ & $J_1 - 5J_2$ \\
		$2$ & $2$ & $0$ & $\pm1$ & $J_1 \mp h$           & $1$ & $2$ & $2$ & $\pm1$ & $J_1 - 5J_2 \mp h$ \\
		$2$ & $2$ & $0$ & $\pm2$ & $J_1 \mp 2h$          & $2$ & $2$ & $2$ & $0$ & $J_1 - 3J_2$ \\
		$2$ & $0$ & $2$ & $0$ & $-2J_1$                  & $2$ & $2$ & $2$ & $\pm1$ & $J_1 - 3J_2 \mp h$ \\
		$2$ & $0$ & $2$ & $\pm1$ & $-2J_1 \mp h$         & $2$ & $2$ & $2$ & $\pm2$ & $J_1 - 3J_2 \mp 2h$ \\
		$2$ & $0$ & $2$ & $\pm2$ & $-2J_1 \mp 2h$        & $3$ & $2$ & $2$ & $0$ & $J_1$ \\
		$1$ & $2$ & $1$ & $0$ & $J_1 -3J_2$              & $3$ & $2$ & $2$ & $\pm1$ & $J_1 \mp h$ \\
		$1$ & $2$ & $1$ & $\pm1$ & $J_1 -3J_2 \mp h$     & $3$ & $2$ & $2$ & $\pm2$ & $J_1 \mp 2h$ \\
		$2$ & $2$ & $1$ & $0$ & $J_1 -J_2$               & $3$ & $2$ & $2$ & $\pm3$ & $J_1 \mp 3h$ \\
		$2$ & $2$ & $1$ & $\pm1$ & $J_1 -J_2 \mp h$      & $4$ & $2$ & $2$ & $0$ & $J_1 + 4J_2$ \\
		$2$ & $2$ & $1$ & $\pm2$ & $J_1 -J_2 \mp 2h$     & $4$ & $2$ & $2$ & $\pm1$ & $J_1 + 4J_2 \mp h$ \\
		$3$ & $2$ & $1$ & $0$ & $J_1 + 2J_2$             & $4$ & $2$ & $2$ & $\pm2$ & $J_1 + 4J_2 \mp 2h$ \\
		$3$ & $2$ & $1$ & $\pm1$ & $J_1 + 2J_2 \mp h$    & $4$ & $2$ & $2$ & $\pm3$ & $J_1 + 4J_2 \mp 3h$ \\
		$3$ & $2$ & $1$ & $\pm2$ & $J_1 + 2J_2 \mp 2h$   & $4$ & $2$ & $2$ & $\pm4$ & $J_1 + 4J_2 \mp 4h$ \\		                                                 
		\hline
	\end{tabular}
\end{table}

At this stage, it is quite straightforward to obtain from the full energy spectrum quoted in Tab.~\ref{t1} an exact result for the partition function of the spin-1 Heisenberg diamond cluster $\mathcal{Z} = \mathrm{Tr}\,{\rm e}^{-\beta \hat{\mathcal{H}}} = \sum_{i=1}^{81} {\rm e}^{-\beta E_i}$ with $\beta  = 1/(k_{\rm B} T)$ ($k_{\rm B}$ is Boltzmann's constant, $T$ is the absolute temperature), which is explicitly given by the following lengthy expression:
\begin{eqnarray}
	\mathcal{Z} &= \mathrm{e}^{-\beta\left(J_1+2J_2\right)}\left[1\!+\!2\cosh{(\beta h)}\!+\!2\cosh{(2\beta h)} \!+\! 2\cosh{(3\beta h)}\right] 
		+ \mathrm{e}^{\beta\left(J_1+2J_2\right)} + \mathrm{e}^{-\beta\left(J_1-6J_2\right)}\nonumber
	\\ &+ \mathrm{e}^{\beta\left(J_1+J_2\right)}\left[2\!+\!4\cosh{(\beta h)}\!+\!2\cosh{(2\beta h)}\right]
		+ \mathrm{e}^{-\beta\left(J_1-J_2\right)}\left[1\!+\!2\cosh{(\beta h)}\!+\!2\cosh{(2\beta h)}\right]\nonumber
	\\ &+ \mathrm{e}^{2\beta J_1}\left[3\!+\!4\cosh{(\beta h)} \!+\! 2\cosh{(2\beta h)}\right]
		+ \mathrm{e}^{-\beta\left(-J_1+J_2\right)}\left[1\!+\!2\cosh{(\beta h)}\!+\!2\cosh{(2\beta h)}\right]\nonumber
	\\ &+ \mathrm{e}^{-\beta\left(-J_1+2J_2\right)}\left[1\!+\!2\cosh{(\beta h)}\!+\!2\cosh{(2\beta h)} \!+\! 2\cosh{(3\beta h)}\right]
		+ \mathrm{e}^{\beta J_1}\left[1\!+\!2\cosh{(\beta h)}\right]\nonumber
	\\ &+ \mathrm{e}^{-\beta J_1}\left[2\!+\!4\cosh{(\beta h)} \!+\! 4\cosh{(2\beta h)} \!+\! 2\cosh{(3\beta h)}\right]
		+ \mathrm{e}^{\beta\left(J_1+3J_2\right)}\left[1\!+\!2\cosh{(\beta h)}\right]\nonumber
	\\ &+ \mathrm{e}^{-\beta\left(J_1-3J_2\right)}\left[2\!+\!4\cosh{(\beta h)}\!+\!2\cosh{(2\beta h)}\right] 
		+ \mathrm{e}^{-\beta\left(J_1-5J_2\right)}\left[1\!+\!2\cosh{(\beta h)}\right]\nonumber
	\\ &+ \mathrm{e}^{-\beta\left(J_1+4J_2\right)}\left[1\!+\!2\cosh{(\beta h)}\!+\!2\cosh{(2\beta h)} \!+\! 2\cosh{(3\beta h)} \!+\! 2\cosh{(4\beta h)}\right].
\label{PF}
\end{eqnarray}
The magnetization per one spin can be subsequently obtained from the associated Gibbs free energy $G = - k_{\rm B} T \ln \mathcal{Z}$ by making use of the following formula:
\begin{eqnarray}
m = - \frac{1}{4} \frac{\partial G}{\partial B} = \frac{g \mu_{\rm B}}{4} \frac{\partial \ln \mathcal{Z}}{\partial(\beta h)} = \frac{g \mu_{\rm B}}{4} \frac{\mathcal{Z}_h}{\mathcal{Z}},
\label{mag}
\end{eqnarray}
whereas the expression $\mathcal{Z}_h \equiv \partial \mathcal{Z}/\partial(\beta h)$ is defined as follows:
\begin{eqnarray}
	\mathcal{Z}_h &= 2 \mathrm{e}^{-\beta\left(J_1+2J_2\right)}\left[\sinh{(\beta h)}\!+\!2\sinh{(2\beta h)} \!+\! 3\sinh{(3\beta h)}\right] \nonumber	\\ 
	   &+ 4 \mathrm{e}^{\beta\left(J_1+J_2\right)}\left[\sinh{(\beta h)}\!+\!\sinh{(2\beta h)}\right]
		+ 2 \mathrm{e}^{-\beta\left(J_1-J_2\right)}\left[\sinh{(\beta h)}\!+\!2\sinh{(2\beta h)}\right]\nonumber 	\\ 
		 &+ 4 \mathrm{e}^{2\beta J_1}\left[\sinh{(\beta h)} \!+\! \sinh{(2\beta h)}\right]
		+ 2 \mathrm{e}^{-\beta\left(-J_1+J_2\right)}\left[\sinh{(\beta h)}\!+\!2\sinh{(2\beta h)}\right]\nonumber 	\\ 
		 &+ 2 \mathrm{e}^{-\beta\left(-J_1+2J_2\right)}\left[\sinh{(\beta h)}\!+\!2\sinh{(2\beta h)} \!+\! 3\sinh{(3\beta h)}\right] 
		+ 2 \mathrm{e}^{\beta J_1} \sinh{(\beta h)}\nonumber 	\\ 
		 &+ 2 \mathrm{e}^{-\beta J_1}\left[2\sinh{(\beta h)} \!+\! 4\sinh{(2\beta h)} \!+\! 3\sinh{(3\beta h)}\right]
		+ 2 \mathrm{e}^{\beta\left(J_1+3J_2\right)} \sinh{(\beta h)} \nonumber \\ 
		 &+ 4 \mathrm{e}^{-\beta\left(J_1-3J_2\right)}\left[\sinh{(\beta h)}\!+\!\sinh{(2\beta h)}\right] 
		+ 2 \mathrm{e}^{-\beta\left(J_1-5J_2\right)}\sinh{(\beta h)}\nonumber 	\\ 
		 &+ 2 \mathrm{e}^{-\beta\left(J_1+4J_2\right)}\left[\sinh{(\beta h)}\!+\!2\sinh{(2\beta h)} \!+\! 3\sinh{(3\beta h)} \!+\! 4\sinh{(4\beta h)}\right].
\label{PFM}
\end{eqnarray}
The magnetic molar entropy of the spin-1 Heisenberg diamond cluster can be similarly obtained from the exact result (\ref{PF}) for the partition function according to the formula:
\begin{eqnarray}
	S_m = - N_{\rm A} \frac{\partial G}{\partial T} = R \left(\ln{\mathcal{Z}} + \frac{T}{\mathcal{Z}}\frac{\partial \mathcal{Z}}{\partial T}\right),
\label{ENT}
\end{eqnarray}
where $N_{\rm A}$ and $R$ stand for Avogadro's and universal gas constant, respectively. It should be mentioned that the final formula for a temperature derivative of the partition function is too lengthy in order to write it down here explicitly.

\section{Theoretical results}
\label{theory}

\begin{figure}
\begin{center}
\includegraphics[width=0.50\textwidth]{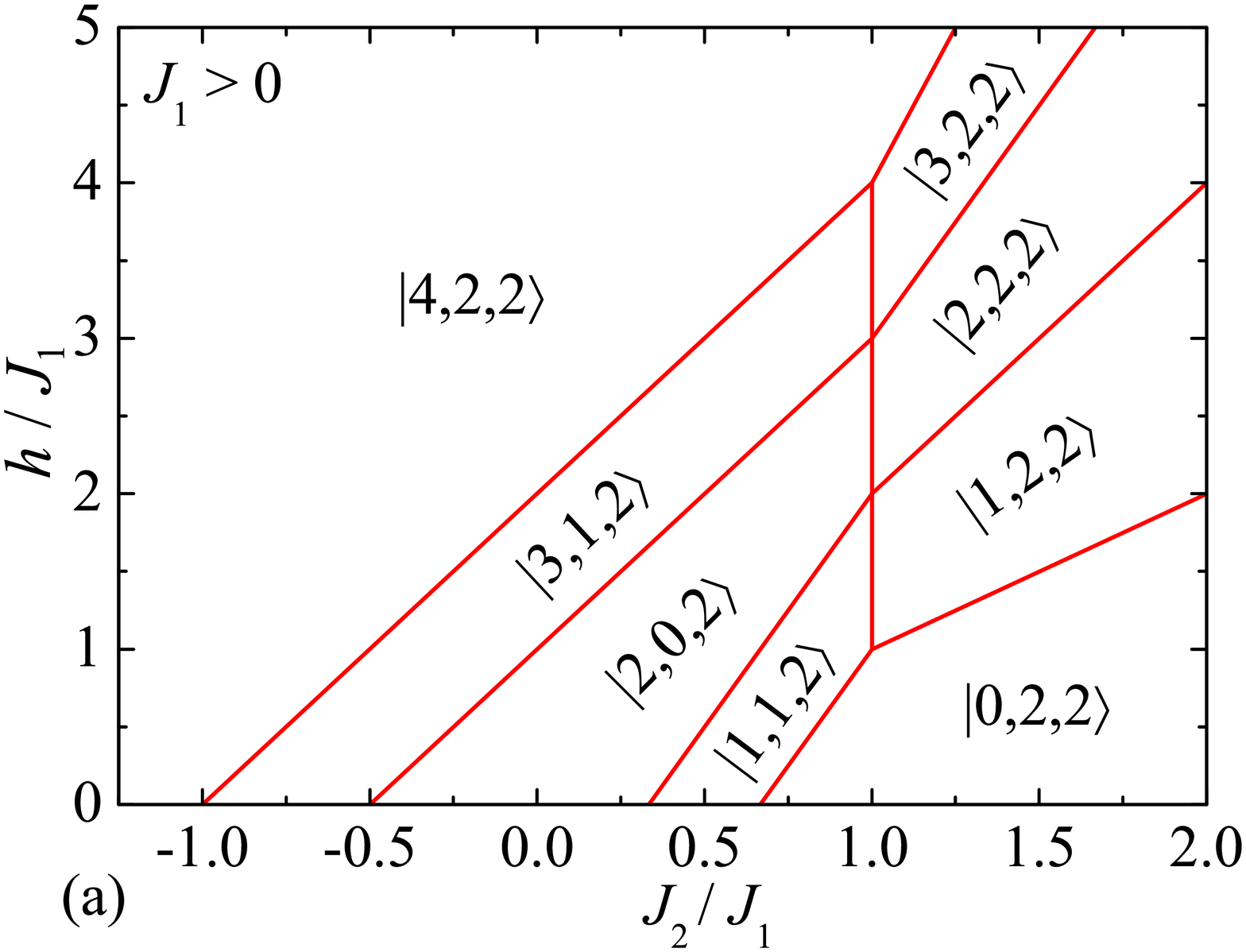}
\hspace{-0.5cm}
\includegraphics[width=0.50\textwidth]{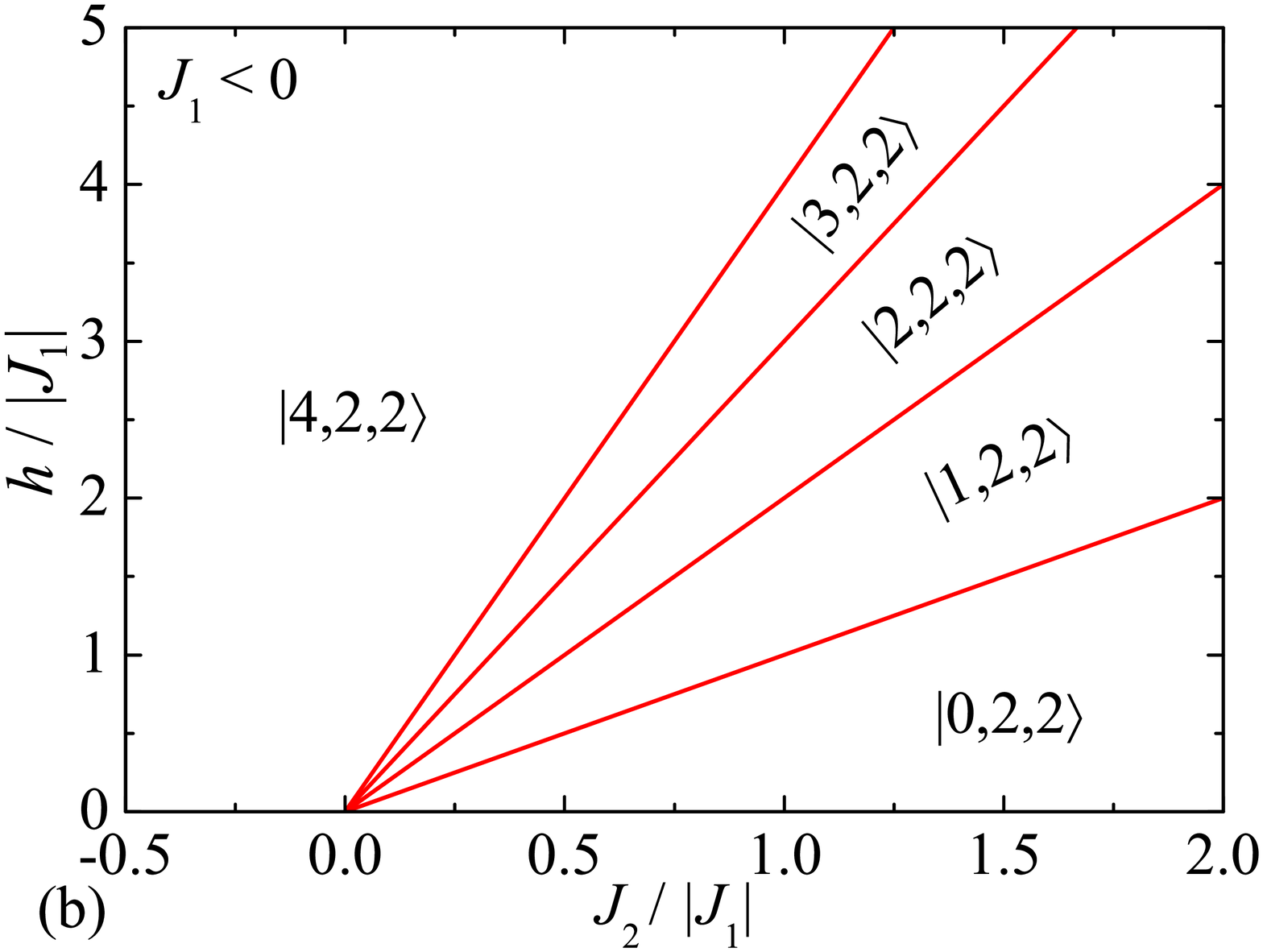}
\end{center}
\vspace{-0.5cm}
\caption{The ground-state phase diagram of the spin-1 Heisenberg diamond cluster in the $J_2/|J_1| - h/|J_1|$ plane for two particular cases with: (a) the antiferromagnetic interaction $J_1>0$; (b) the ferromagnetic interaction $J_1<0$. The eigenvectors $|S_T=S_T^z,S_{12},S_{34}\rangle$ are specified according to the quantum spin numbers determining the total spin and its $z$-component $S_T=S_T^z$, as well as, two composite spins $S_{12}$ and $S_{34}$ formed by spin-1 entities from opposite corners of a diamond spin cluster.}
\label{fig2}       
\end{figure}

In this part, we will proceed to a comprehensive analysis of the most interesting results for the ground state, magnetization curves and magnetocaloric properties of the spin-1 Heisenberg diamond cluster. The ground-state phase diagram of the spin-1 Heisenberg diamond cluster is displayed in Fig.~\ref{fig2} in the $J_2/|J_1|-h/|J_1|$ plane for two particular cases, which differ from one another in antiferromagnetic ($J_1>0$) vs. ferromagnetic ($J_1<0$) character of the coupling constant along a shorter diagonal of the diamond spin cluster. One finds by inspection eight different ground states unambiguously given by the eigenvectors $|S_T=S_T^z,S_{12},S_{34}\rangle$, which are classified through a set of the quantum spin numbers determining the total spin and its $z$-component being equal $S_T=S_T^z$ within all ground states, as well as, two composite spins $S_{12}$ and $S_{34}$ formed by spin-1 entities from opposite corners of the diamond spin cluster. Within the framework of the Kambe's coupling scheme \cite{kamb50,sinn70}, it is convenient to express first the relevant ground states as a linear combination over a tensor product of eigenvectors of two considered spin pairs $\vert S_T, S_{12}, S_{34}\rangle = \sum _{i} a_i \vert S_{12}, S_{12}^z \rangle \otimes \vert S_{34}, S_{34}^z \rangle$ before writing them more explicitly as a linear combination over spin states of the usual Ising basis $\vert S_T, S_{12}, S_{34}\rangle = \sum _{i} b_i \vert S_{1}^z, S_{2}^z, S_{3}^z, S_{4}^z \rangle$. The exact formulas for the eigenvectors $\vert S_{12}, S_{12}^z \rangle$ and $\vert S_{34}, S_{34}^z \rangle$ of the spin-1 Heisenberg dimers are not quoted here explicitly, because they can be found in our preceding work \cite{stre05}.

\begin{figure}
\begin{center}
\includegraphics[width=0.7\textwidth]{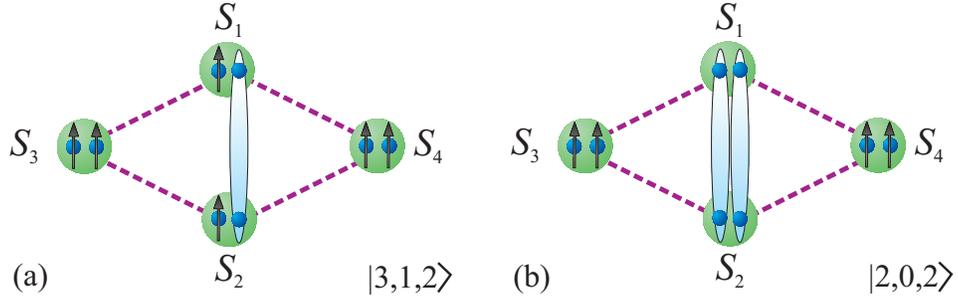}
\end{center}
\vspace{-0.5cm}
\caption{A schematic representation of two valence-bond-crystal ground states given by the eigenvectors: (a) $\vert 3, 1, 2 \rangle$; (b) $\vert 2, 0, 2 \rangle$. Each spin-1 particle (large green sphere) is symmetrically decomposed within the valence-bond-solid picture into two spin-1/2 entities (small blue spheres), which either form a singlet (valence) bond schematically demarcated by ovals or are polarized into the magnetic-field direction as specified by up-pointing arrows.}
\label{vbc}       
\end{figure}

The spin-1 Heisenberg diamond cluster of course shows at high enough magnetic fields a classical ferromagnetic state with fully saturated magnetic moment of all four constituent spins:
\begin{equation}
\vert 4, 2, 2 \rangle = \vert{2, 2}\rangle \!\otimes\! \vert{2, 2}\rangle = \vert 1, 1, 1, 1 \rangle,
\label{eq:422}
\end{equation}
which changes below the saturation field to one of two eigenvectors $\vert 3, 1, 2 \rangle$ or $\vert 3, 2, 2 \rangle$ with character of the one-magnon deviation from the fully polarized ferromagnetic state:
\begin{eqnarray}
\vert 3, 1, 2 \rangle \!\!\!&=&\!\!\! \vert{1, 1}\rangle \!\otimes\! \vert{2, 2}\rangle = \frac{1}{\sqrt{2}}\left(\vert 1, 0, 1, 1\rangle \!-\! \vert 0, 1, 1, 1\rangle \right), 
\label{eq:312} \\
\vert 3, 2, 2 \rangle \!\!\!&=&\!\!\! \frac{1}{\sqrt{2}}(\vert{2, 1}\rangle \!\otimes\! \vert{2, 2}\rangle \!-\! \vert{2, 2}\rangle \!\otimes\! \vert{2, 1}\rangle) \nonumber \\
	 \!\!\!&=&\!\!\! \frac{1}{2}(\vert 1, 1, 1, 0 \rangle \!+\! \vert 1, 1, 0, 1\rangle \!-\! \vert 1, 0, 1, 1\rangle \!-\! \vert 0, 1, 1, 1\rangle). \label{eq:322}
\end{eqnarray}
Although the ground states $\vert 3, 1, 2 \rangle$ and $\vert 3, 2, 2 \rangle$ essentially lead in a zero-temperature magnetization curve to the intermediate 3/4-plateau with the same value of the total magnetization, it is quite clear from Eqs. (\ref{eq:312}) and (\ref{eq:322}) that the underlying mechanism for formation of the relevant intermediate magnetization plateau is very different. While in the former ground state $\vert 3, 1, 2 \rangle$ the further-distant spins $\langle S_3^z \rangle = \langle S_4^z \rangle = 1$ from the 'wings' of the butterfly tetramer contribute to the total magnetization twice as large as the near-distant counterparts $\langle S_1^z \rangle = \langle S_2^z \rangle = 1/2$ from its 'main body', all four spins contribute equally to the total magnetization $\langle S_1^z \rangle = \langle S_2^z \rangle = \langle S_3^z \rangle = \langle S_4^z \rangle = 3/4$ in the latter ground state $\vert 3, 2, 2 \rangle$. It could be thus concluded that the spin density is homogeneously distributed over the whole diamond spin cluster in the eigenstate $\vert 3, 2, 2 \rangle$ what is in sharp contrast with the eigenstate $\vert 3, 1, 2 \rangle$, which has character of the valence-bond crystal with a singlet bond formed within the near-distant spin pair (see Fig. \ref{vbc}(a) for a schematic illustration). The similar situation is encountered also in other two eigenvectors $\vert 2, 0, 2 \rangle$ and $\vert 2, 2, 2 \rangle$:   	
\begin{eqnarray}
\vert 2, 0, 2 \rangle \!\!\!&=&\!\!\! \vert{0, 0}\rangle \!\otimes\! \vert{2, 2}\rangle= \frac{1}{\sqrt{3}}\left(\vert 1, \!-\!1, 1, 1 \rangle \!+\! \vert \!-\!1, 1, 1, 1 \rangle \!-\! \vert 0, 0, 1, 1 \rangle\right), \label{eq:202} \\
\vert 2, 2, 2 \rangle \!\!\!&=&\!\!\! \sqrt{\frac{2}{7}}(\vert{2, 2}\rangle \!\otimes\! \vert{2, 0}\rangle \!+\! \vert{2, 0}\rangle \!\otimes\! \vert{2, 2}\rangle) \!-\! \sqrt{\frac{3}{7}}(\vert{2, 1}\rangle \!\otimes\! \vert{2, 1}\rangle) \nonumber \\ 
	 \!\!\!&=&\!\!\! \sqrt{\frac{2}{7}}\frac{1}{\sqrt{6}}(\vert \!-\!1, 1, 1, 1 \rangle \!+\! \vert 1, \!-\!1, 1, 1\rangle \!+\! \vert 1, 1, \!-\!1, 1 \rangle \!+\! \vert 1, 1, 1, \!-\!1 \rangle \!+\!2 \vert 1, 1, 0, 0 \rangle \nonumber \\ 
	 \!\!\!&+&\!\!\! 2\vert 0, 0, 1, 1 \rangle) \!-\! \sqrt{\frac{3}{7}}\frac{1}{2}(\vert 1, 0, 1, 0 \rangle \!+\! \vert 1, 0, 0, 1 \rangle \!+\! \vert 0, 1, 1, 0 \rangle \!+\! \vert 0, 1, 0, 1 \rangle), \label{eq:222}
\end{eqnarray}
which may eventually become the respective ground state at lower magnetic fields being responsible for emergence of the intermediate 1/2-plateau in a zero-temperature magnetization curve. The former ground state $\vert 2, 0, 2 \rangle$ can be regarded as another type of a valence-bond crystal with an inhomogeneous distribution of the spin density; the near-distant spins $\langle S_1^z \rangle = \langle S_2^z \rangle = 0$ do not contribute anyhow to the total magnetization due to formation of double-singlet bonds in contrast with the further-distant spins $\langle S_3^z \rangle = \langle S_4^z \rangle = 1$ providing the highest possible contribution owing to their fully polarized nature (see Fig. \ref{vbc}(b) for a schematic illustration). Contrary to this, the ground state $\vert 2, 2, 2 \rangle$ can be characterized through a homogeneous distribution of the spin density $\langle S_1^z \rangle = \langle S_2^z \rangle = \langle S_3^z \rangle = \langle S_4^z \rangle = 1/2$ spanned over the whole diamond spin cluster. The inhomogeneous vs. homogeneous distribution of the spin density can be also encountered in other two ground states $\vert 1, 1, 2 \rangle$ and $\vert 1, 2, 2 \rangle$:
\begin{eqnarray}
\vert 1, 1, 2 \rangle \!\!\!&=&\!\!\! -\frac{1}{\sqrt{10}}(\vert{1, 1}\rangle \!\otimes\! \vert{2, 0}\rangle) \!+\! \sqrt{\frac{3}{10}}\left(\vert{1, 0}\rangle \!\otimes\! \vert{2, 1}\rangle\right)\!+\! \sqrt{\frac{6}{10}}(\vert {1, -1} \rangle \!\otimes\! \vert {2, 2}\rangle) \nonumber \\
	 \!\!\!&=&\!\!\! -\frac{1}{\sqrt{10}}\frac{1}{\sqrt{12}}(\vert 1, 0, 1, \!-\!1 \rangle \!+\! \vert 1, 0, \!-\!1, 1\rangle \!+\! 2\vert 1, 0, 0, 0 \rangle \!-\! \vert 0, 1, 1, \!-\!1 \rangle \nonumber \\
	\!\!\!&-&\!\!\! \vert 0, 1, \!-\!1, 1 \rangle \!-\! 2\vert 0, 1, 0, 0 \rangle) \!+\! \sqrt{\frac{3}{10}}\frac{1}{2}(\vert 1, \!-\!1, 1, 0 \rangle \!+\! \vert 1, \!-\!1, 0, 1 \rangle \nonumber \\
	 \!\!\!&-&\!\!\! \vert \!-\!1, 1, 1, 0 \rangle \!-\! \vert \!-\!1, 1, 0, 1 \rangle) \!+\! \sqrt{\frac{6}{10}}\frac{1}{\sqrt{2}}(\vert \!-\!1, 0, 1, 1\rangle \!-\! \vert 0, \!-\!1, 1, 1 \rangle), 
\label{eq:112} \\
\vert 1, 2, 2 \rangle \!\!\!&=&\!\!\! \frac{1}{\sqrt{5}}(\vert{2, 2}\rangle \!\otimes\! \vert{2, -1}\rangle \!-\! \vert{2, -1}\rangle \!\otimes\! \vert{2, 2}\rangle) \!-\! \sqrt{\frac{3}{10}}(\vert{2, 1}\rangle \!\otimes\! \vert{2, 0}\rangle \!+\! \vert{2, 0}\rangle \!\otimes\! \vert{2, 1}\rangle) \nonumber \\
	  \!\!\!&=&\!\!\! \frac{1}{\sqrt{10}}(\vert 1, 1, \!-\!1, 0 \rangle \!+\! \vert 1, 1, 0, \!-\!1 \rangle \!-\! \vert \!-\!1, 0, 1, 1 \rangle \!-\! \vert 0, \!-\!1, 1, 1 \rangle) \nonumber \\
	  \!\!\!&-&\!\!\! \sqrt{\frac{3}{10}}\frac{1}{\sqrt{12}}(\vert 1, 0, 1, \!-\!1 \rangle \!+\! \vert 1, 0, \!-\!1, 1 \rangle \!+\! 2\vert 1, 0, 0, 0\rangle \!+\! \vert 0, 1, 1, \!-\!1 \rangle \!+\! \vert 0, 1, \!-\!1, 1 \rangle \nonumber \\
	  \!\!\!&+&\!\!\! 2\vert 0, 1, 0, 0 \rangle) \!+\! \sqrt{\frac{3}{10}}\frac{1}{\sqrt{12}}(\vert 1, \!-\!1, 1, 0 \rangle \!+\! \vert 1, \!-\!1, 0, 1 \rangle \!+\! \vert \!-\!1, 1, 1, 0\rangle \!+\! \vert \!-\!1, 1, 0, 1 \rangle \nonumber \\
	 \!\!\!&+&\!\!\! 2\vert 0, 0, 1, 0 \rangle \!+\! 2\vert 0, 0, 0, 1 \rangle), \label{eq:122}
\end{eqnarray}
which are responsible for appearance of the intermediate 1/4-plateau in a zero-temperature magnetization curve. Although the former ground state $\vert 1, 1, 2 \rangle$ cannot be classified as a valence-bond-crystal state, the spin density is inhomogeneously distributed within this eigenstate as convincingly evidenced by an opposite sign of the local magnetizations of the near- and further-distant spin pairs $\langle S_1^z \rangle = \langle S_2^z \rangle = -1/4$ and $\langle S_3^z \rangle = \langle S_4^z \rangle = 3/4$, respectively. On the contrary, the spin density is homogeneously distributed over the whole diamond spin cluster in the latter ground state $\vert 1, 2, 2 \rangle$, which has identical local magnetizations of all four spins  $\langle S_1^z \rangle = \langle S_2^z \rangle = \langle S_3^z \rangle = \langle S_4^z \rangle = 1/4$. Finally, the last ground state $\vert 0, 2, 2 \rangle$ can be characterized by the eigenvector: 
\begin{eqnarray}
\vert 0, 2, 2 \rangle \!\!\!&=&\!\!\! \frac{1}{\sqrt{5}}(\vert{2, 2}\rangle \!\otimes\! \vert{2, -2}\rangle \!+\! \vert{2, -2}\rangle \!\otimes\! \vert{2, 2}\rangle) \!-\! \frac{1}{\sqrt{5}}(\vert{2, 1}\rangle \!\otimes\! \vert{2, -1}\rangle \!+\! \vert{2, -1}\rangle \!\otimes\! \vert{2, 1}\rangle) \nonumber \\
	  \!\!\!&+&\!\!\! \frac{1}{\sqrt{5}}(\vert{2, 0}\rangle \!\otimes\! \vert{2, 0}\rangle) \!=\! \frac{1}{\sqrt{5}}(\vert 1, 1, \!-\!1, \!-\!1 \rangle \!+\! \vert \!-\!1, \!-\!1, 1, 1\rangle) \!-\! \frac{1}{\sqrt{5}}\frac{1}{2}(\vert 1, 0, \!-\!1, 0 \rangle \nonumber \\
	  \!\!\!&+&\!\!\! \vert 1, 0, 0, \!-\!1 \rangle \!+\! \vert 0, 1, \!-\!1, 0 \rangle \!+\! \vert 0, 1, 0, \!-\!1\rangle \!+\! \vert \!-\!1, 0, 1, 0 \rangle \!+\! \vert \!-\!1, 0, 0, 1 \rangle \!+\! \vert 0, \!-\!1, 0, 1 \rangle \nonumber \\
	  \!\!\!&+&\!\!\! \vert 0, \!-\!1, 1, 0 \rangle) \!+\! \frac{1}{\sqrt{5}}\frac{1}{6}(\vert 1, \!-\!1, 1, \!-\!1 \rangle \!+\! \vert 1, \!-\!1, \!-\!1, 1 \rangle \!+\! 2\vert 1, \!-\!1, 0, 0 \rangle \!+\! \vert \!-\!1, 1, 1, \!-\!1 \rangle \nonumber \\
	  \!\!\!&+&\!\!\! \vert \!-\!1, 1, \!-\!1, 1 \rangle \!+\! 2 \vert \!-\!1, 1, 0, 0 \rangle \!+\! 2\vert 0, 0, 1, \!-\!1 \rangle \!+\! 2 \vert 0, 0, \!-\!1, 1 \rangle \!+\! 4\vert 0, 0, 0, 0 \rangle),
\label{eq:022}
\end{eqnarray}
which implies a complete absence of the local magnetization for all four constituent spins $\langle S_1^z \rangle = \langle S_2^z \rangle = \langle S_3^z \rangle = \langle S_4^z \rangle = 0$ and is thus responsible for the onset of zero magnetization plateau. 

It is worthwhile to remark that the critical magnetic fields, which determine a magnetic-field-driven transition between two ground states with a homogeneous distribution of the spin density, can be obtained according to the following formula:
\begin{itemize}
\item $|n-1,2,2\rangle \to |n,2,2\rangle$: $h_{c,n}= n J_2$ \quad for \quad $n = 1,2,3,4$,
\end{itemize}
whereas the critical magnetic fields determining a phase coexistence of the ground states with an inhomogeneous distribution of the spin density are explicitly given by:  
\begin{itemize}
\item $|3,1,2\rangle \to |4,2,2\rangle$: $h_{c,5}=2J_2+2J_1$,
\item $|2,0,2\rangle \to |3,1,2\rangle$: $h_{c,6}=2J_2+J_1$,
\item $|1,1,2\rangle \to |2,0,2\rangle$: $h_{c,7}=3J_2-J_1$,
\item $|0,2,2\rangle \to |1,1,2\rangle$: $h_{c,8}=3J_2-2J_1$.
\end{itemize}
It should be also mentioned that the ground states $|n,2,2\rangle$ ($n=0,1,2,3,4$) with the homogeneous spin density are realized predominantly in the parameter region where the antiferromagnetic coupling constant along sides of a diamond spin cluster overwhelms over the antiferromagnetic coupling constant along its shorter diagonal $J_2/J_1 > 1, J_1 >0$ [see Fig.~\ref{fig2}(a)] or alternatively the coupling constant along the shorter diagonal becomes ferromagnetic $J_1<0$ [see Fig.~\ref{fig2}(b)]. 

The ground-state phase diagrams displayed in Fig.~\ref{fig2} also shed light on a diversity of zero-temperature magnetization curves. It follows from Fig.~\ref{fig2}(a) that the spin-1 Heisenberg diamond cluster with the antiferromagnetic coupling constant $J_1>0$ along sides of a diamond spin cluster exhibits six different magnetization profiles depending on a relative strength of the coupling constants $J_2/J_1$. The zero-temperature magnetization curve of the spin-1 Heisenberg diamond cluster should accordingly reflect four field-induced transitions $|0,2,2\rangle \to |1,2,2\rangle \to |2,2,2\rangle \to |3,2,2\rangle \to |4,2,2\rangle$ for $J_2/J_1 > 1$, other four field-driven transitions $|0,2,2\rangle \to |1,1,2\rangle \to |2,0,2\rangle \to |3,1,2\rangle \to |4,2,2\rangle$ for $J_2/J_1 \in (2/3, 1)$, three field-induced transitions $|1,1,2\rangle \to |2,0,2\rangle \to |3,1,2\rangle \to |4,2,2\rangle$ for $J_2/J_1 \in (1/3, 2/3)$, two field-driven transitions $|2,0,2\rangle \to |3,1,2\rangle \to |4,2,2\rangle$ for $J_2/J_1 \in (-1/2, 1/3)$, the single field-induced transition $|3,1,2\rangle \to |4,2,2\rangle$ for $J_2/J_1 \in (-1, -1/2)$ or is without any field-driven transition for $J_2/J_1 <-1$. To compare with, the zero-temperature magnetization process of the spin-1 Heisenberg diamond cluster with the ferromagnetic coupling constant $J_1<0$ along sides of a diamond spin cluster is much less diverse, since it either shows a sequence of four field-induced transitions $|0,2,2\rangle \to |1,2,2\rangle \to |2,2,2\rangle \to |3,2,2\rangle \to |4,2,2\rangle$ for $J_2 > 0$ or is without any field-driven transition for $J_2 < 0$. Since both these magnetization scenarios can be also found in the former case with the antiferromagnetic coupling constant $J_1>0$ we will henceforth restrict our attention to this particular case.
 
\begin{figure}[h!]
\includegraphics[width=0.5\textwidth]{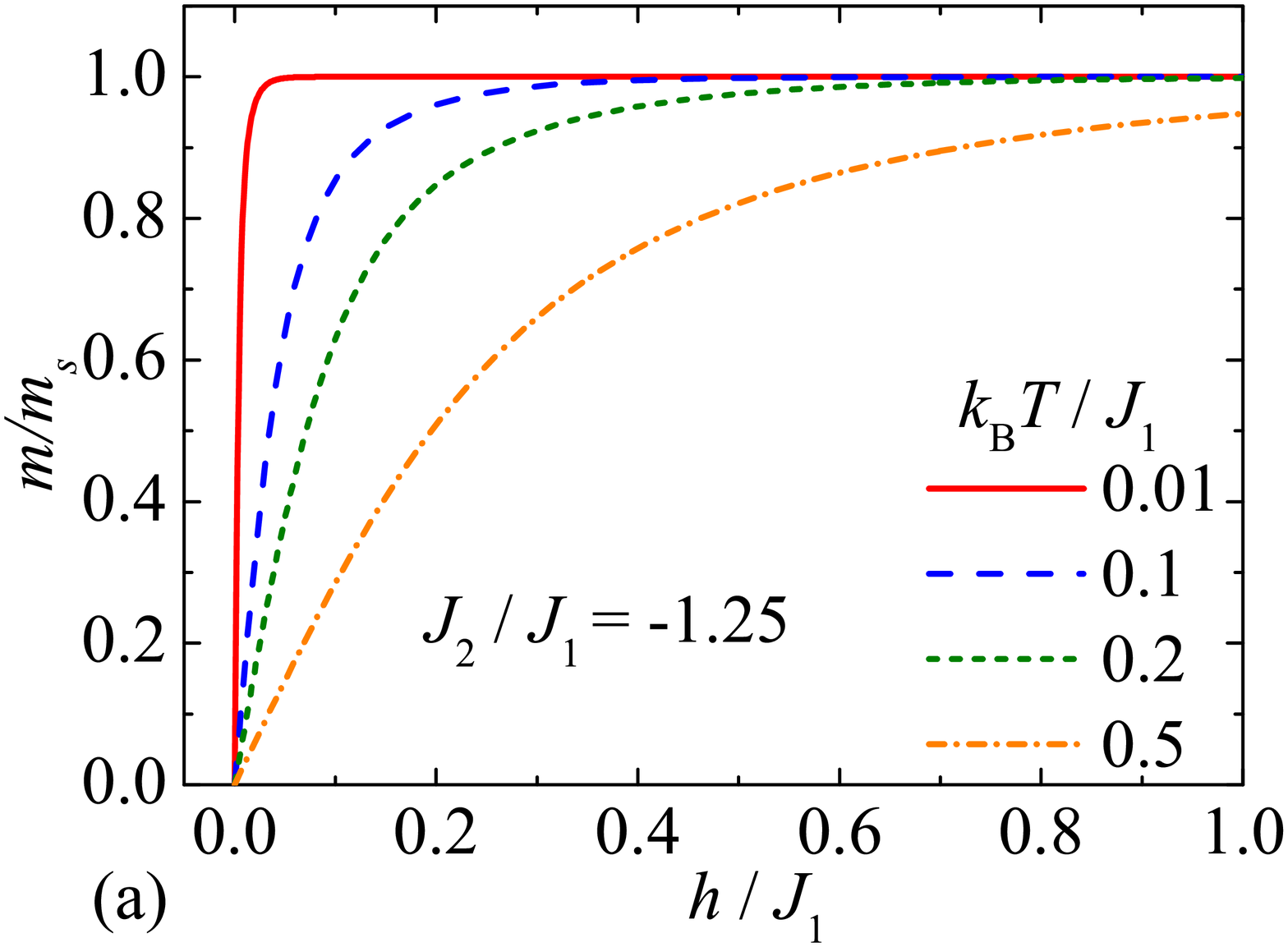}
\hspace{-0.7cm}
\includegraphics[width=0.5\textwidth]{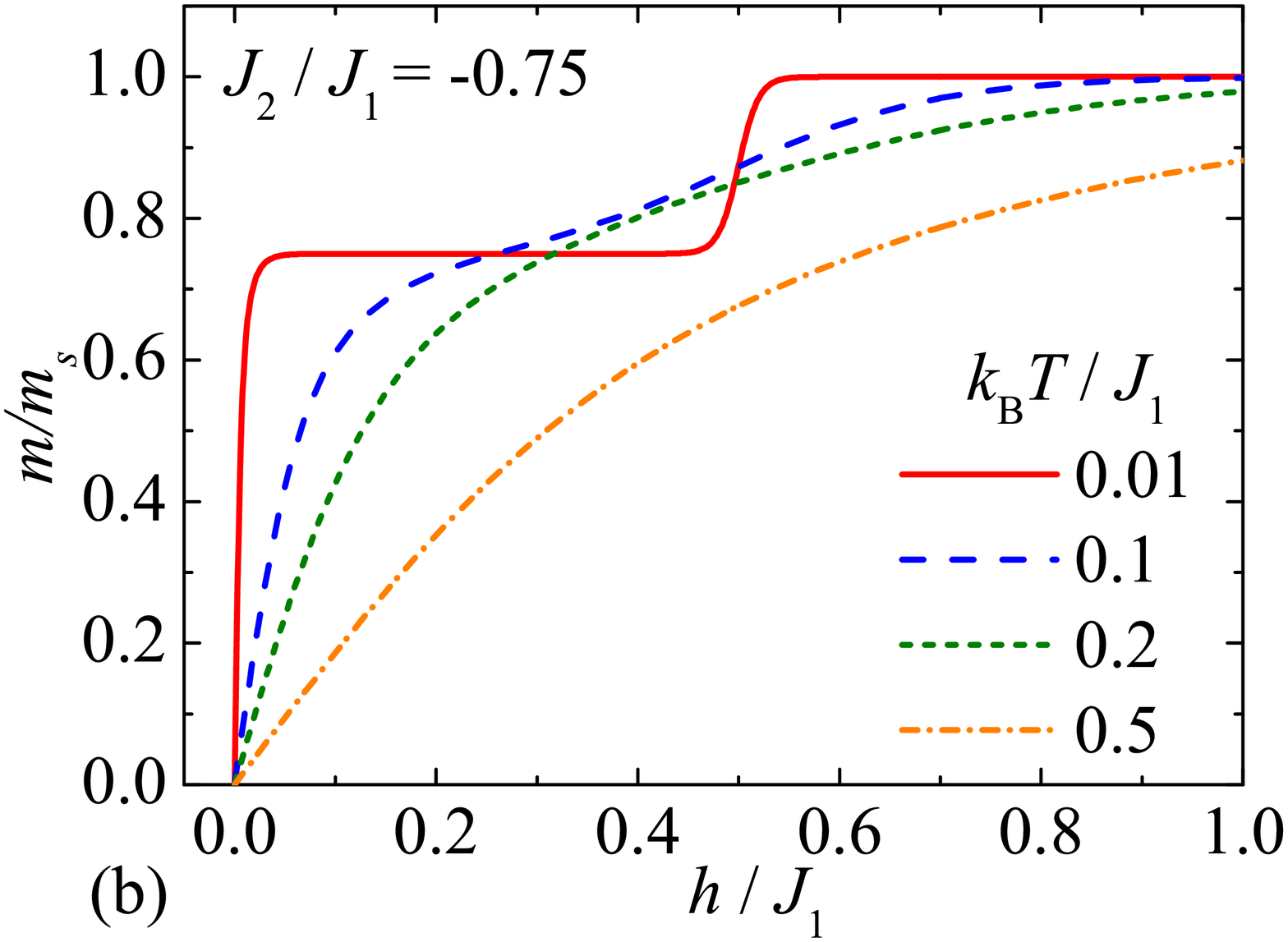}
\hspace{-0.7cm}
\includegraphics[width=0.5\textwidth]{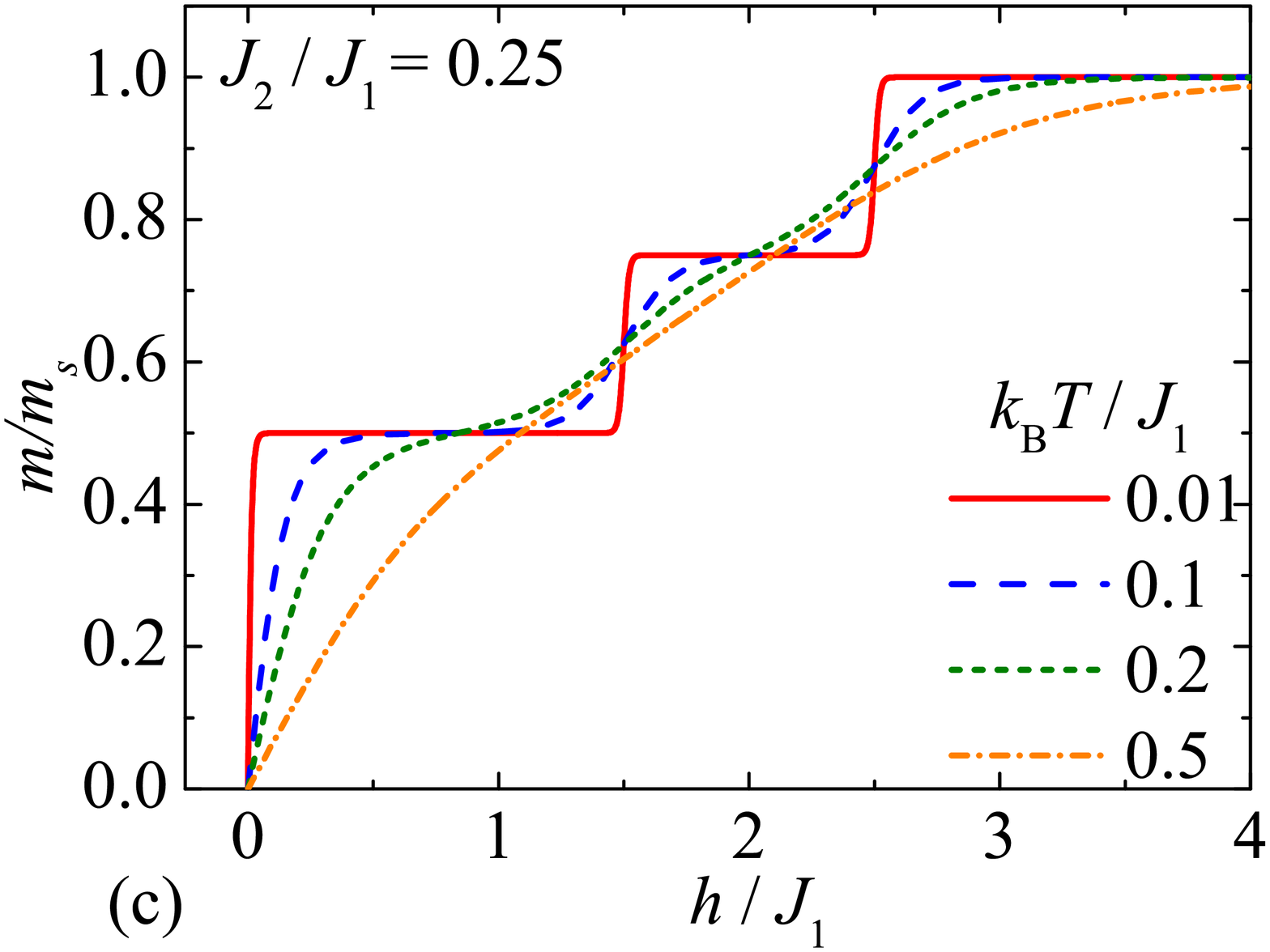}
\hspace{-0.7cm}
\includegraphics[width=0.5\textwidth]{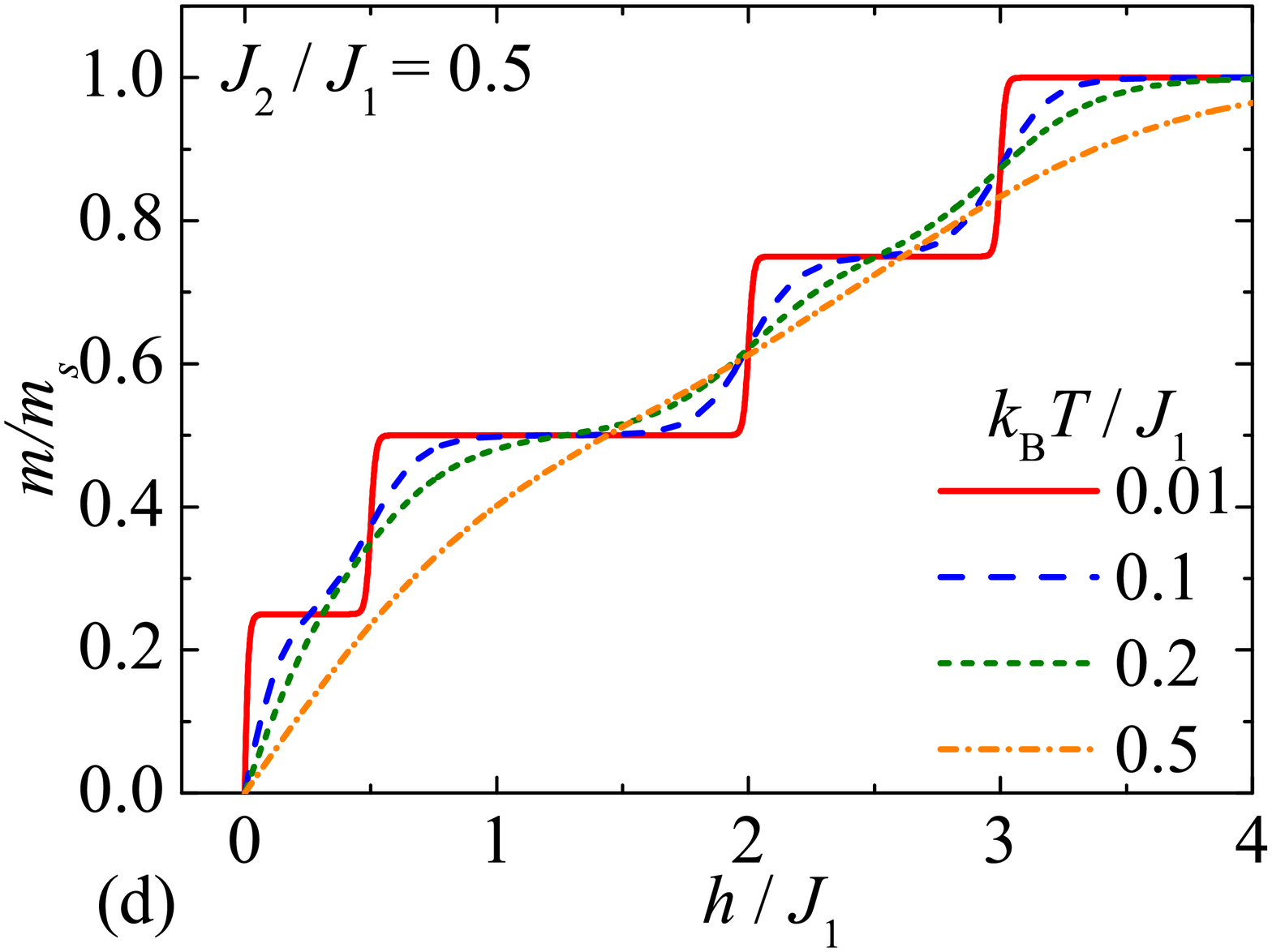}
\hspace{-0.7cm}
\includegraphics[width=0.5\textwidth]{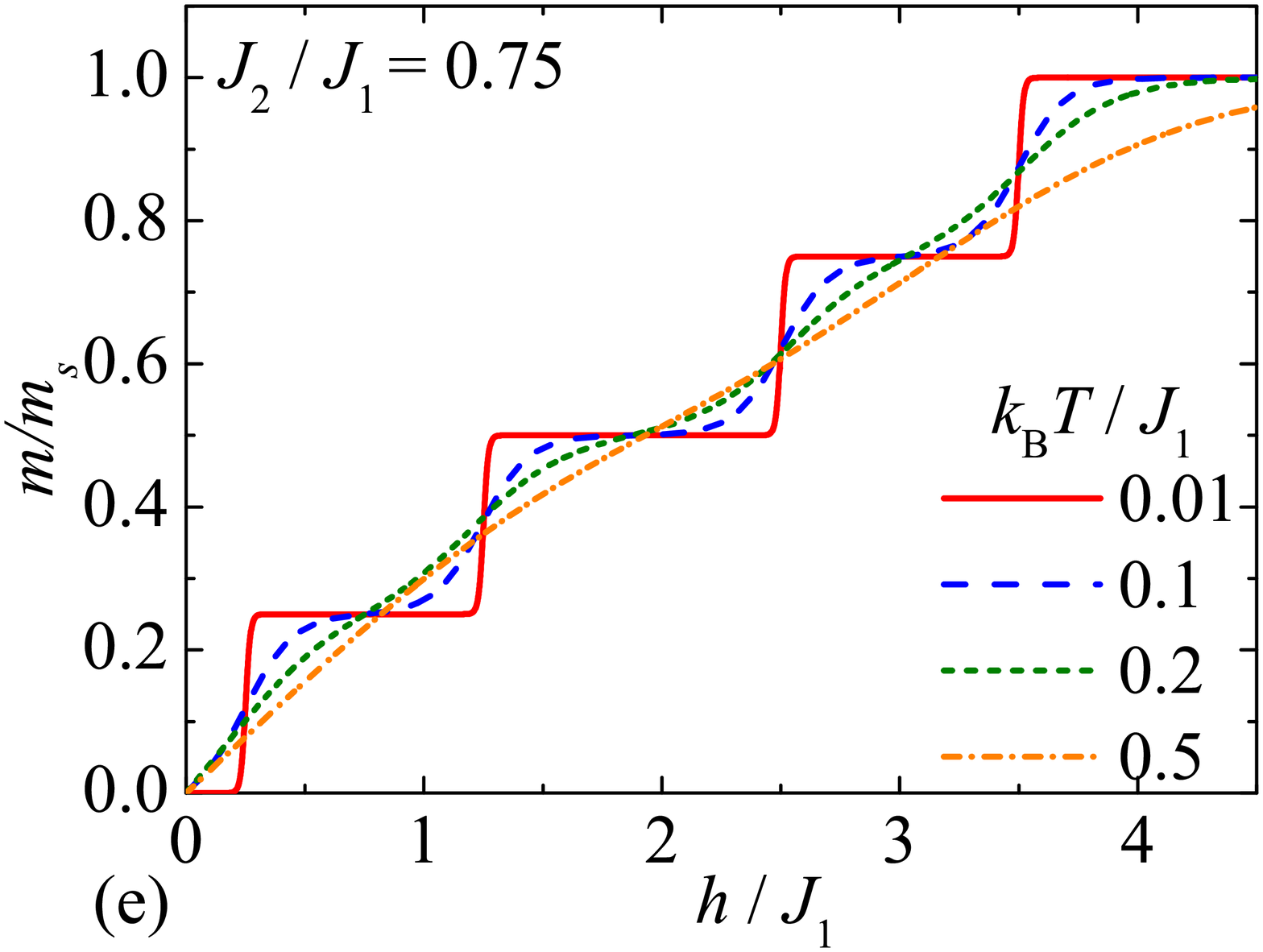}
\hspace{-0.2cm}
\includegraphics[width=0.5\textwidth]{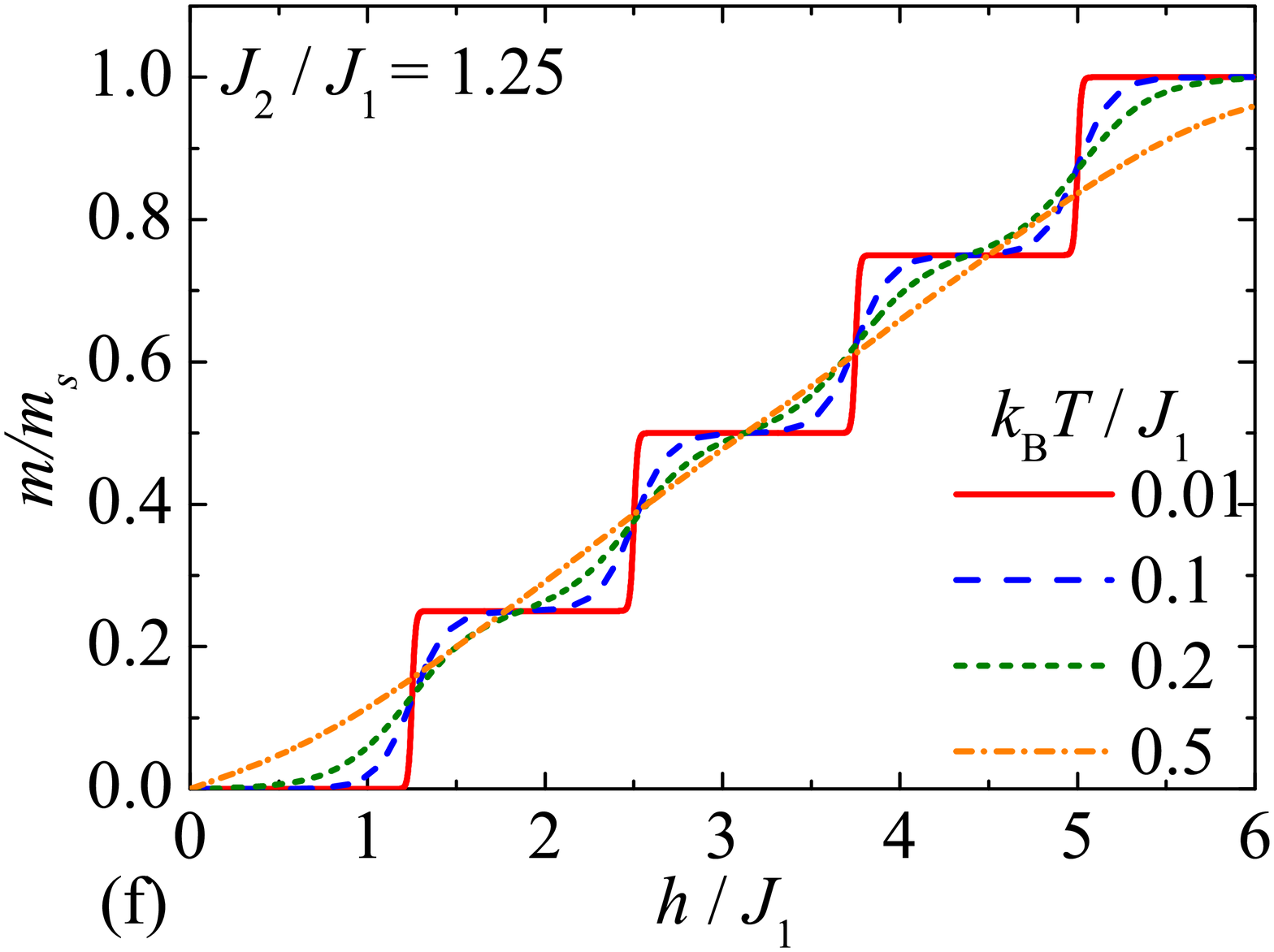}
\caption{Isothermal magnetization curves of the spin-1 Heisenberg diamond cluster with the antiferromagnetic coupling constant $J_1 > 0$ along its shorter diagonal for four different values of temperature and a few selected values of the interaction ratio: (a) $J_2/J_1 = -1.25$; (b) $J_2/J_1 = -0.75$; (c) $J_2/J_1 = 0.25$; (d) $J_2/J_1 = 0.5$; (e) $J_2/J_1 = 0.75$; (f) $J_2/J_1 = 1.25$. The magnetization is normalized with respect to its saturation value $m_s$.}
\label{figm}       
\end{figure}

Typical isothermal magnetization curves of the spin-1 Heisenberg diamond cluster are plotted in Fig.~\ref{figm} for the antiferromagnetic interaction $J_1>0$ and a few selected values of the interaction ratio $J_2/J_1$ in order to provide an independent check of all possible magnetization profiles and field-driven phase transitions. It should be emphasized that the magnetization curves calculated at the lowest temperature $k_{\rm{B}}T/J_1=0.01$ are strongly reminiscent of zero-temperature magnetization curves with discontinuous jumps of the magnetization, which take place at the aforementioned critical magnetic fields in agreement with the ground-state phase diagram shown in Fig.~\ref{fig2}(a). Note furthermore that the rising temperature causes just a gradual melting of the relevant magnetization curves. The first particular case, which is shown in Fig.~\ref{figm}(a) for the interaction ratio $J_2/J_1 = -1.25$ with the dominant ferromagnetic interaction along the sides of a diamond spin cluster, illustrates a smooth magnetization curve without any intermediate plateau. The second particular case with the weaker ferromagnetic interaction $J_2/J_1 = -0.75$ shows an abrupt rise of the magnetization in vicinity of zero magnetic field, which is subsequently followed by the intermediate 3/4-plateau ending up just at the saturation field [see Fig.~\ref{figm}(b)]. It is noteworthy that the intermediate 3/4-plateau as well as a steep rise of the magnetization close to the saturation field is gradually smeared out upon increasing of temperature. The magnetization curves with a steep rise of the magnetization followed by the intermediate 1/2- and 3/4-plateaus is depicted in Fig.~\ref{figm}(c) for the specific value of the interaction ratio $J_2/J_1=0.25$.  The magnetization curves of the spin-1 Heisenberg diamond cluster displayed in Fig.~\ref{figm}(d) for the higher value of the interaction ratio $J_2/J_1=0.5$ indicate presence of the intermediate 1/4-, 1/2- and 3/4-plateaus, which follow-up the initial abrupt rise of the magnetization observable near zero magnetic field. It should be stressed, moreover, that the most narrow 1/4-plateau becomes already indiscernible at relatively low temperature $k_{\rm{B}}T/J_1 \approx 0.1$ due to its tiny energy gap. The magnetization curves of the spin-1 Heisenberg diamond cluster for the last two values of the interaction ratio $J_2/J_1=0.75$ and $1.25$, which are plotted in Fig.~\ref{figm}(e) and (f), respectively, imply existence of four intermediate plateaus at 0, 1/4, 1/2 and 3/4 of the saturation magnetization. In spite of their similarity, the underlying mechanism for formation of the magnetization plateaus is preserved just for zero plateau, while the microscopic nature of all other magnetization plateaus is completely different as evidenced by the ground-state phase diagram shown in Fig.~\ref{fig2}(a). 

\begin{figure}
\vspace{-1cm}
\begin{center}
\includegraphics[width=0.45\textwidth]{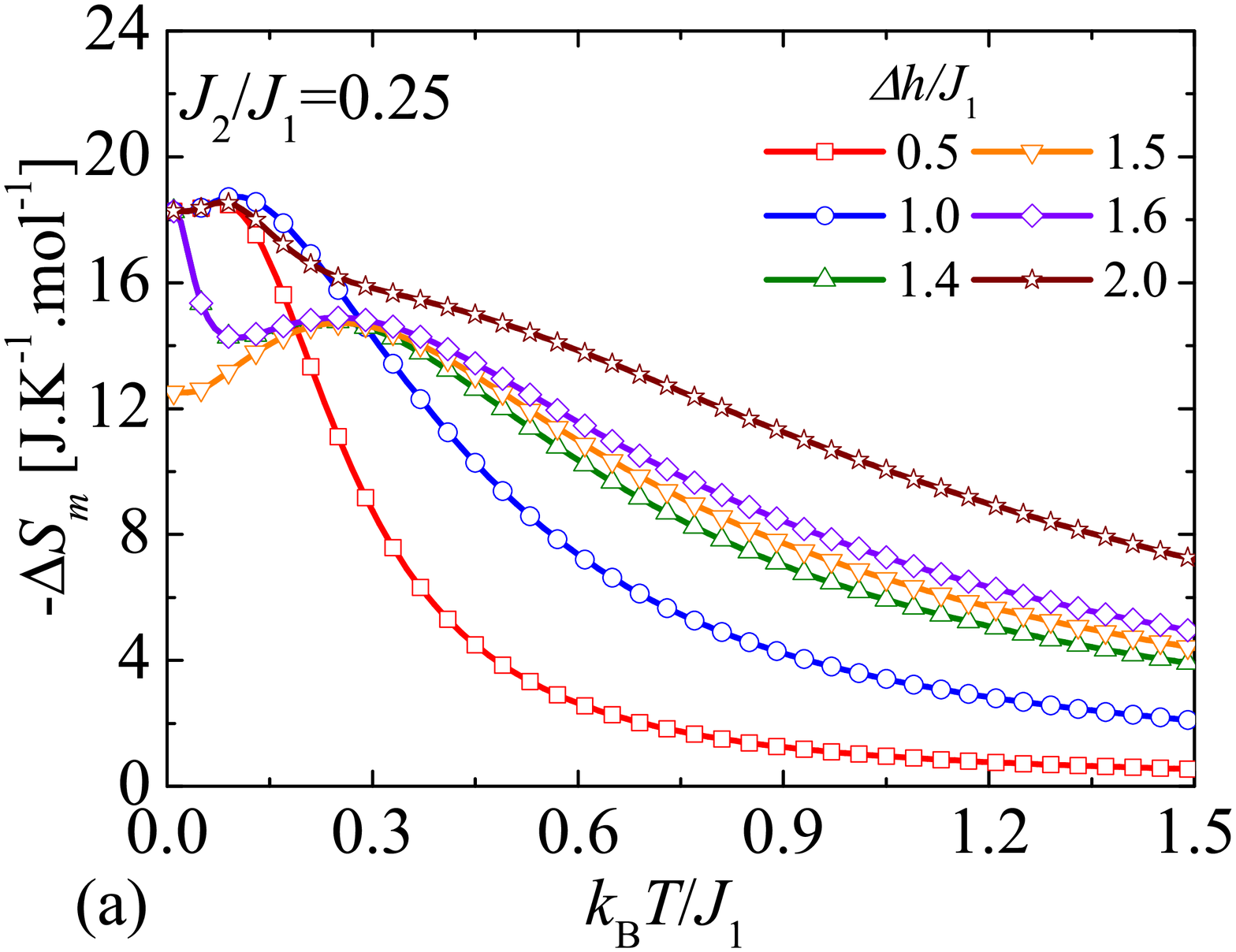}
\hspace{-0.7cm}
\includegraphics[width=0.45\textwidth]{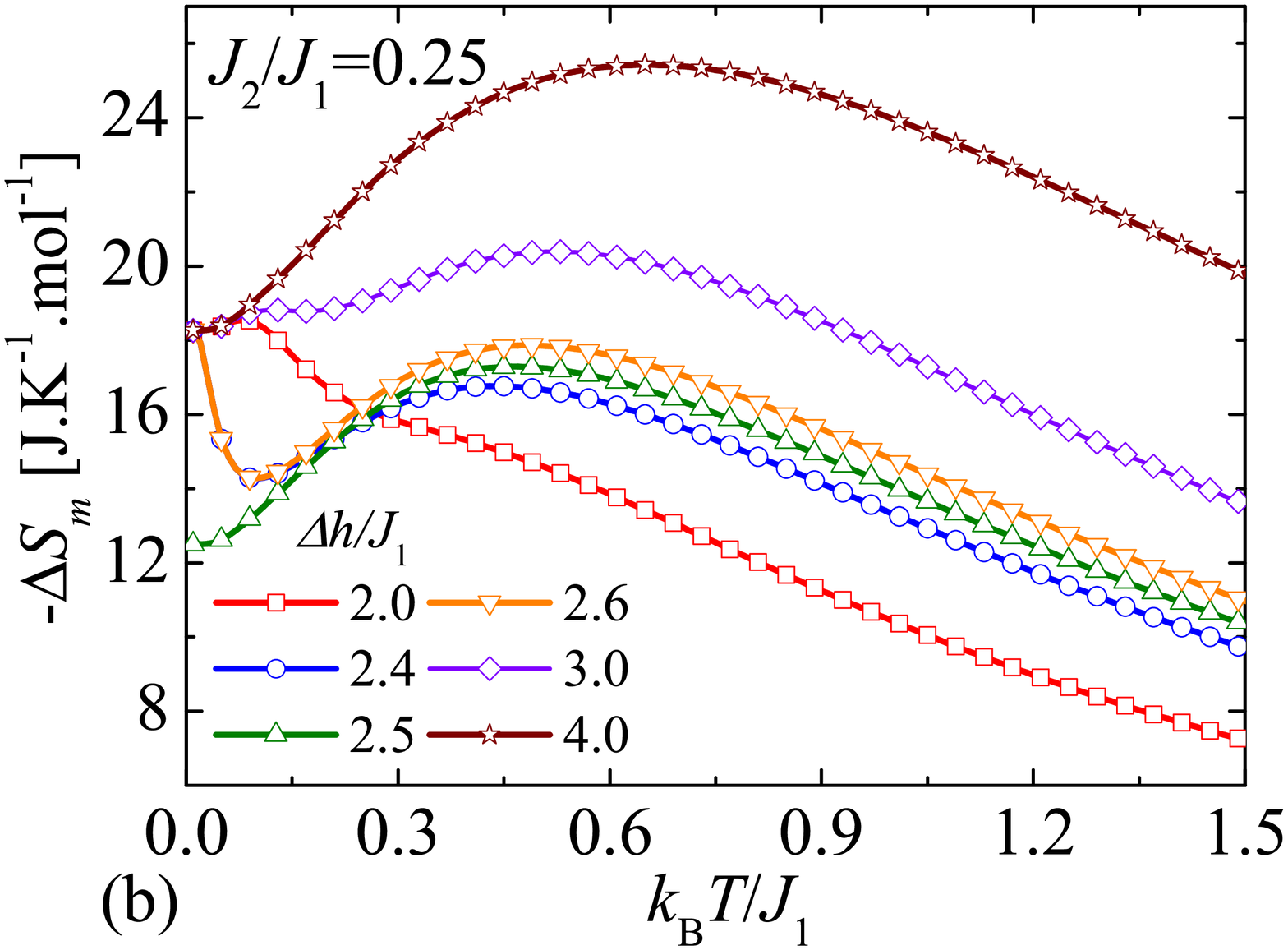}
\hspace{-0.7cm}
\includegraphics[width=0.45\textwidth]{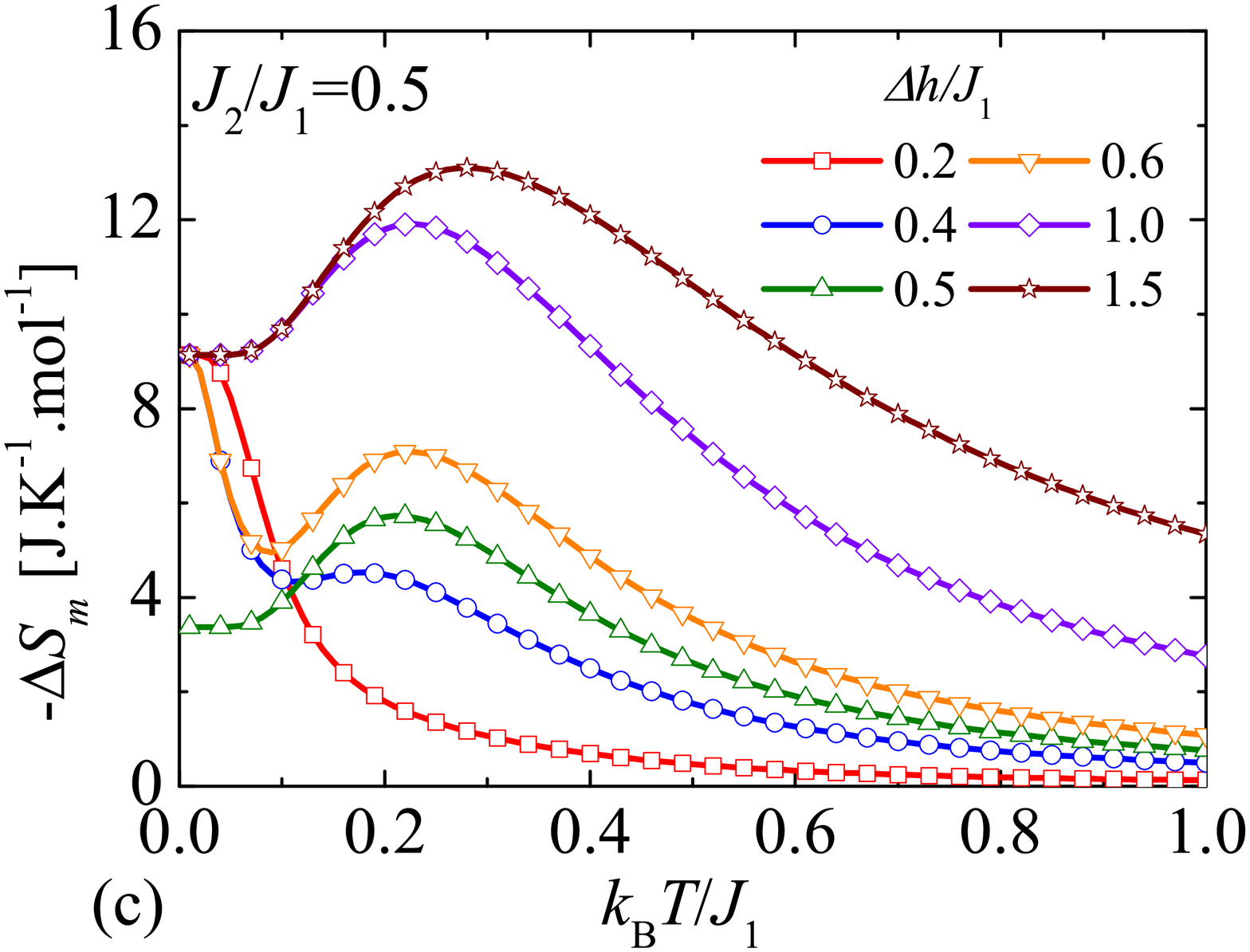}
\hspace{-0.7cm}
\includegraphics[width=0.45\textwidth]{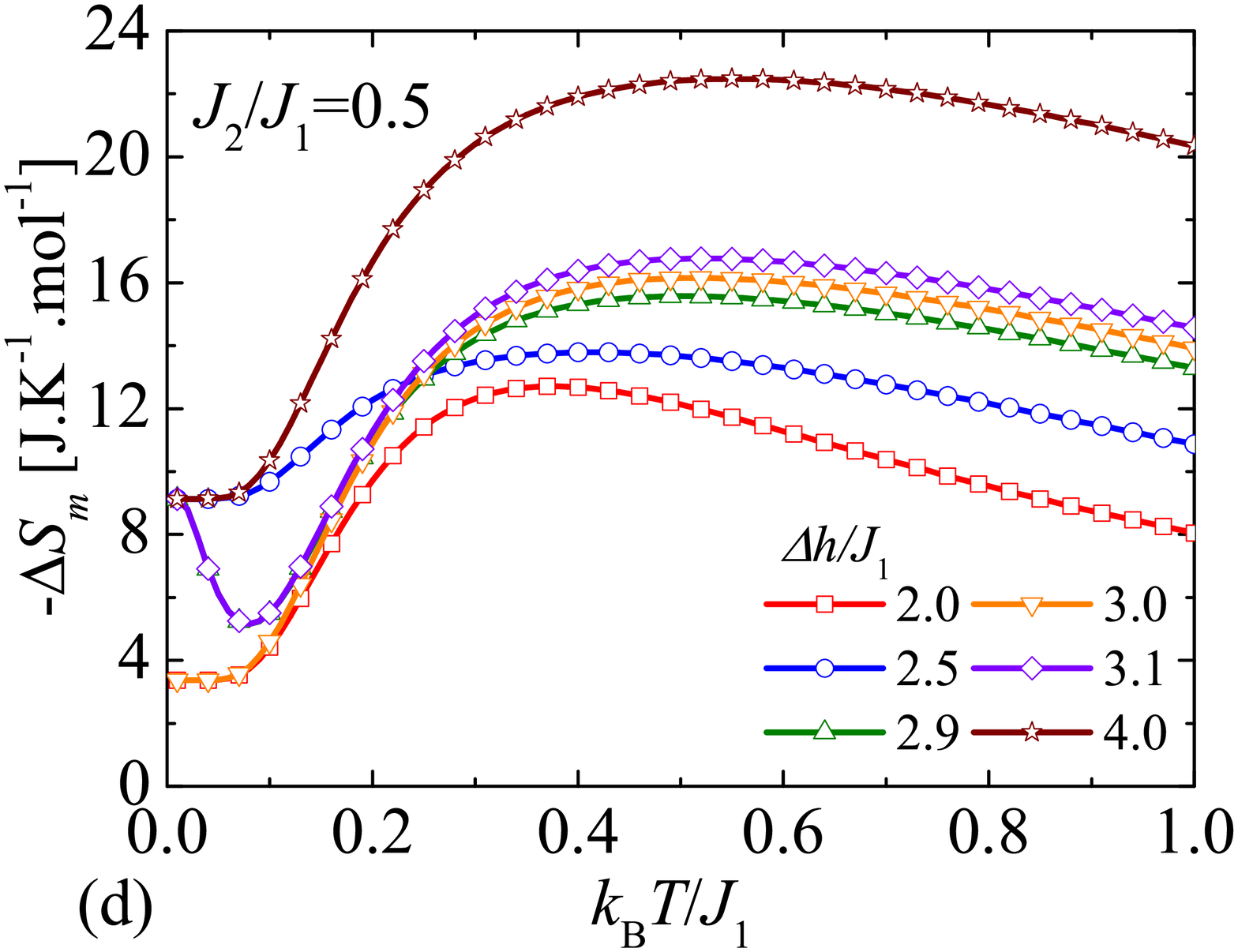}
\hspace{-0.7cm}
\includegraphics[width=0.45\textwidth]{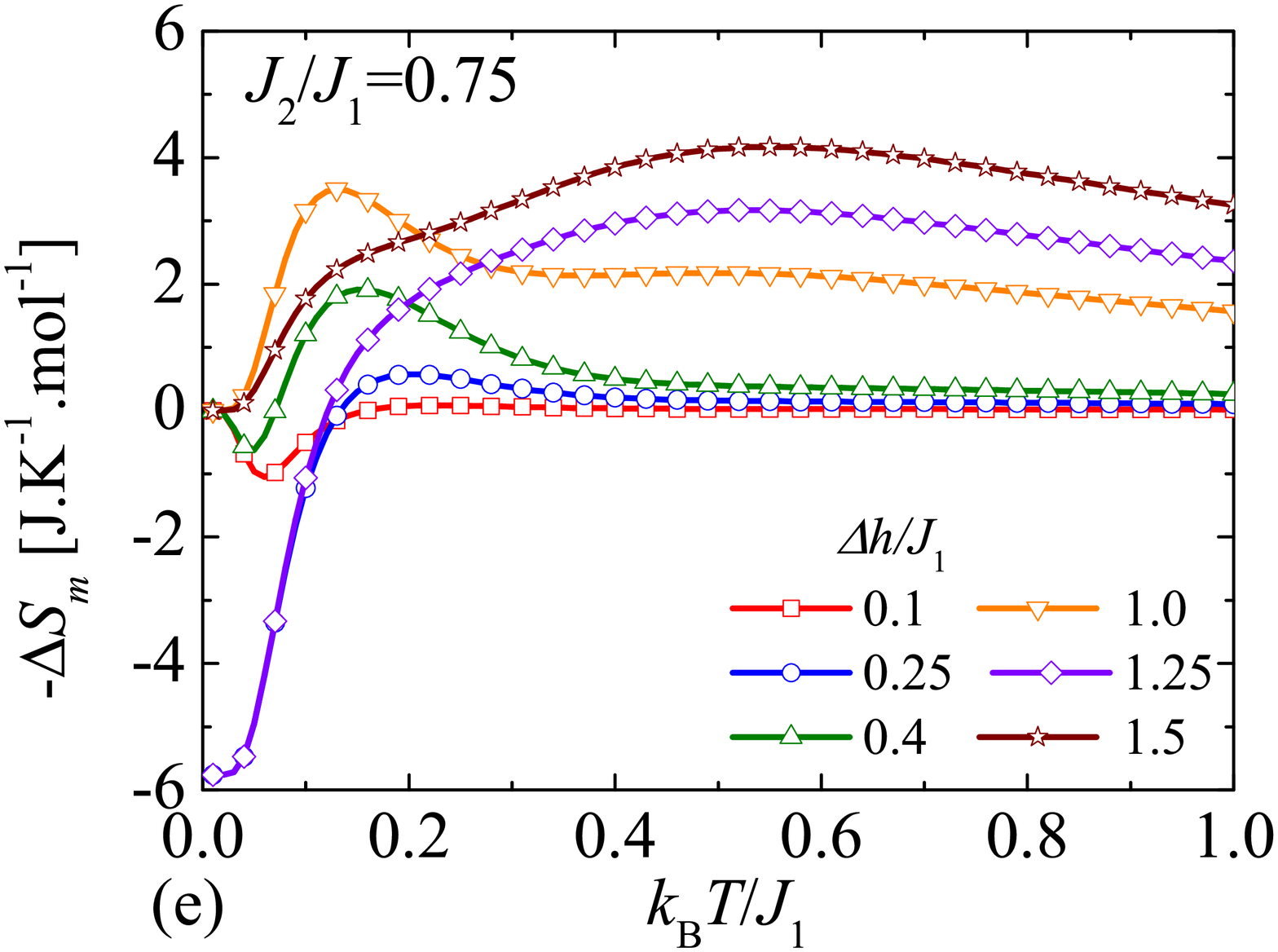}
\hspace{-0.7cm}
\includegraphics[width=0.45\textwidth]{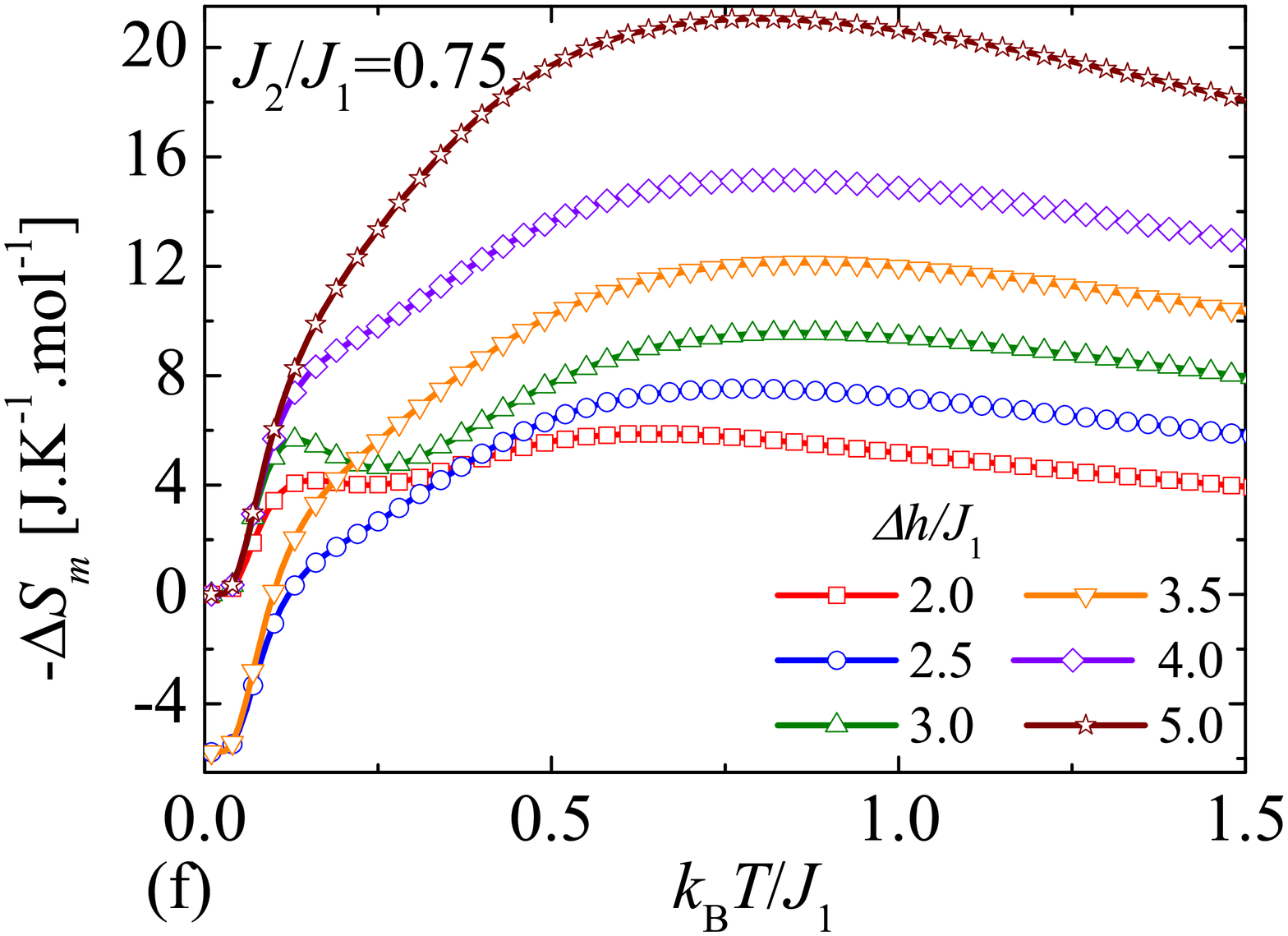}
\hspace{-0.7cm}
\includegraphics[width=0.45\textwidth]{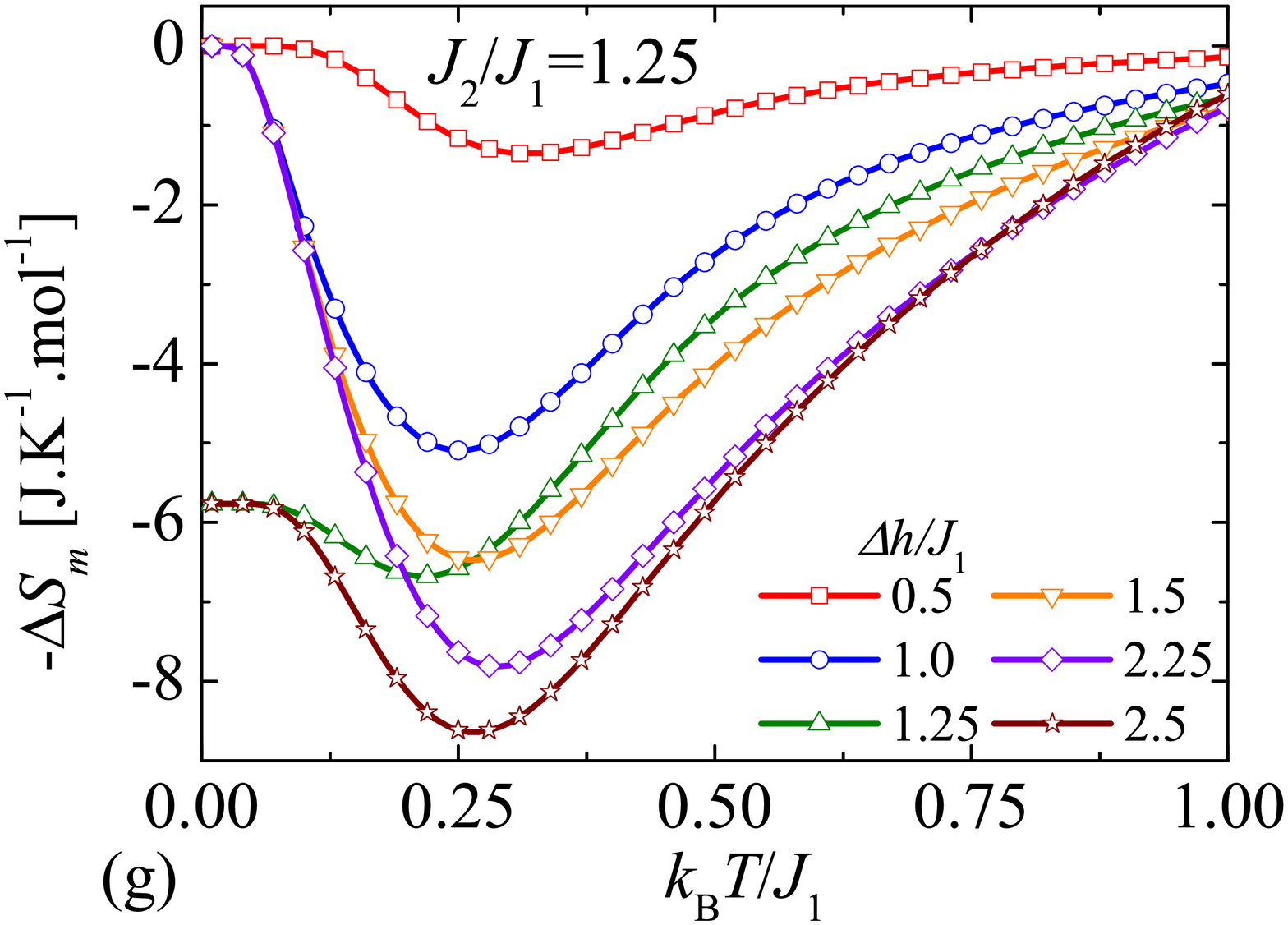}
\hspace{-0.3cm}
\includegraphics[width=0.45\textwidth]{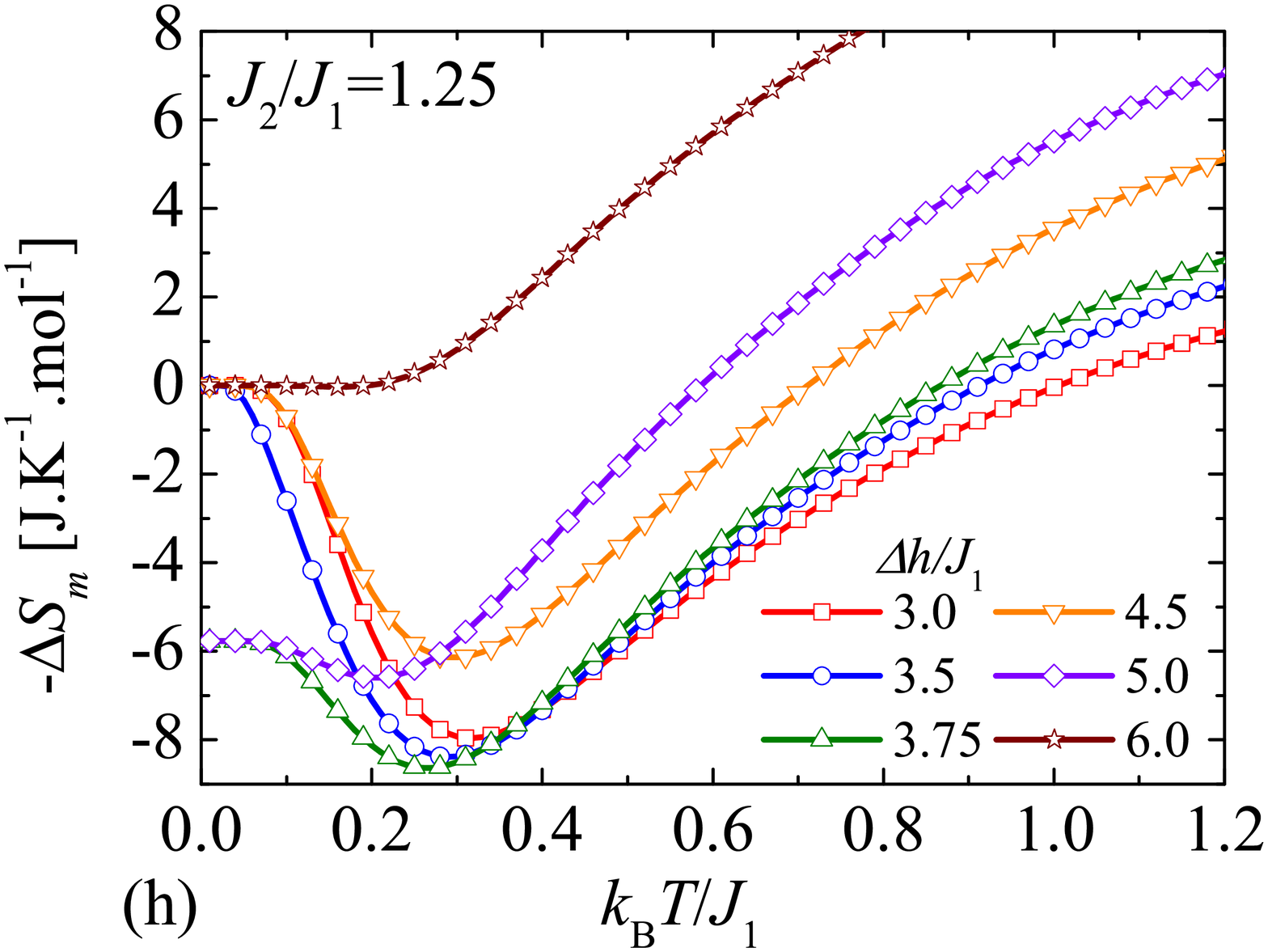}
\end{center}
\vspace{-1cm}
\caption{Temperature variations of the isothermal molar entropy change in J.K$^{-1}$.mol$^{-1}$ units for several values of the magnetic-field change $\Delta h/J_1$ and four different values of the interaction ratio: (a)-(b) $J_2/J_1 = 0.25$; (c)-(d) $J_2/J_1 = 0.5$; (e)-(f) $J_2/J_1 = 0.75$; (g)-(h) $J_2/J_1 = 1.25$.}
\label{fig4}       
\end{figure}

The isothermal entropy change of the spin-1 Heisenberg diamond cluster invoked by the change of magnetic field $\Delta h = h_i - h_f$ is plotted in Fig.~\ref{fig4} as a function of temperature for four different values of the interaction ratio $J_2/J_1$, whereas $h_i \neq 0$ stands for the initial magnetic field and $h_f = 0$ is the final magnetic field during the isothermal demagnetization. Within the proposed notation the conventional MCE occurs for positive values of the isothermal entropy change $-\Delta S_m= S_m(h_f=0)-S_m(h_i\neq 0) >0$, while the inverse MCE is manifested through its negative values $-\Delta S_m<0$.  It should be pointed out, moreover, that the zero-temperature asymptotic value of the molar entropy change $-\Delta S_m = R \ln \Omega_0$ can be simply related to a degeneracy $\Omega_0$ of the zero-field ground state whenever the magnetic-field change does not coincide with any critical magnetic field $\Delta h \neq h_{c,n}$. In the reverse case $\Delta h = h_{c,n}$ the molar entropy change converges in the zero-temperature limit to the smaller asymptotic value $-\Delta S_m = R (\ln \Omega_0 - \ln 2)$ due to a two-fold degeneracy of two coexistent ground states at a critical magnetic field $h_{c,n}$.

The temperature dependences of the molar entropy change of the spin-1 Heisenberg diamond cluster is shown in Fig.~\ref{fig4}(a) and (b) for a few different values of the magnetic-field change and the fixed value of the interaction ratio $J_2/J_1=0.25$, which is consistent with presence of the valence-bond-crystal ground state $|2,0,2\rangle$ in the zero-field limit. It is worthwhile to remark that the singlet state of the near-distant spin pair emergent within the ground state $|2,0,2\rangle$ effectively decouples all spin correlations of two further-distant spins. Owing to this fact, the further-distant spins behave at zero magnetic field as free paramagnetic entities and the respective degeneracy of the zero-field ground-state is $\Omega_0 = 9$. It can be seen from Fig.~\ref{fig4}(a) and (b) that the molar entropy change actually tends to the specific value $-\Delta S_m=R\ln 9\approx 18.3$ J.K$^{-1}$.mol$^{-1}$ for all magnetic-field changes except those being equal to the critical magnetic fields $\Delta h/J_1=1.5$ and 2.5. In this latter case, the molar entropy change acquires in zero-temperature limit smaller asymptotic value $-\Delta S_m=R(\ln 9-\ln 2)\approx 12.5$ J.K$^{-1}$.mol$^{-1}$ in accordance with the previous argumentation [see the curves for $\Delta h/J_1=1.5$ and 2.5 in Fig.~\ref{fig4}(a) and (b)]. Although the isothermal entropy change generally diminishes upon increasing of temperature, it is quite evident from Fig.~\ref{fig4}(a) and (b) that the reverse may be true in a range of moderate temperatures whenever the magnetic-field change is chosen sufficiently close to one of the critical magnetic fields.

The isothermal entropy changes of the spin-1 Heisenberg diamond cluster are depicted in Fig.~\ref{fig4}(c) and (d) for relatively small and moderate changes of the magnetic field by assuming the interaction ratio $J_2/J_1=0.5$ supporting another zero-field ground state $|1,1,2\rangle$. It should be pointed out that the conventional MCE with $-\Delta S_m>0$ occurs for any magnetic-field change quite similarly as in the previous case. In spite of this qualitative similarity, the molar entropy change converges in zero-temperature limit to completely different asymptotic values on account of a triply degenerate ($\Omega_0 = 3$) ground state $|1,1,2\rangle$ realized in zero-field limit. As a matter of fact, it is obvious from Fig. \ref{fig4}(c) and (d) that the molar entropy change reaches either the asymptotic value $-\Delta S_m= R\ln 3\approx 9.1$ J.K$^{-1}$.mol$^{-1}$ or $-\Delta S_m=R(\ln 3-\ln 2)\approx 3.4$ J.K$^{-1}$.mol$^{-1}$ depending on whether or not the magnetic-field change coincides with the critical magnetic field, whereas the latter smaller value of $-\Delta S_m$ 
applies only if the magnetic-field change corresponds to one of three critical magnetic fields $\Delta h/J_1=0.5$, 2.0 or 3.0. Under these specific conditions, the isothermal entropy change starts from this lower asymptotic value, then it increases with rising temperature to its local maximum before it finally tends to zero upon further increase of temperature. The most interesting temperature dependences of the isothermal entropy change can be found when the magnetic-field change is selected slightly below or above the critical magnetic fields [e.g. $\Delta h/J_1=0.4$ or 0.6 in Fig. \ref{fig4}(c)], because the molar entropy change then starts from its higher zero-temperature asymptotic limit, then it shows a rapid decline to a local minimum subsequently followed by a continuous rise to a local maximum upon increasing of temperature before it finally decays to zero in the high-temperature region. 
If the magnetic-field change is sufficiently far from the critical magnetic fields one either finds a monotonic temperature decline of the isothermal entropy change upon increasing of temperature [see curve for $\Delta h/J_1=0.2$ in Fig. \ref{fig4}(c)] or one recovers a nonmonotonic temperature dependence with a single round maximum emerging at some moderate temperature [see the curves for $\Delta h/J_1=1.0$ and 1.5 in Fig. \ref{fig4}(c) or $\Delta h/J_1=4.0$ in Fig. \ref{fig4}(d)]. 

The completely different magnetocaloric features of the spin-1 Heisenberg diamond cluster can be traced back from temperature variations of the isothermal entropy change, which are shown in Fig. \ref{fig4}(e)-(h) for two selected values of the interaction ratio $J_2/J_1=0.75$ and 1.25. The common feature of these two particular cases is that the zero-field ground state is the non-degenerate singlet state $|0,2,2\rangle$, which is responsible for existence of zero magnetization plateau in the respective low-temperature magnetization curves [see Fig.~\ref{figm}(e) and (f)]. in the consequence of that, the molar entropy change asymptotically tends in zero-temperature limit either to zero or to the specific value $-\Delta S_m=-R\ln 2\approx -5.8$ J.K$^{-1}$.mol$^{-1}$ depending on whether the magnetic-field change differs or equals to the critical magnetic fields, respectively. It can be seen from Fig. \ref{fig4}(g) and (h) that the spin-1 Heisenberg diamond cluster with the interaction ratio $J_2/J_1=1.25$ exhibits the inverse MCE with $-\Delta S_m<0$ for most of the magnetic-field changes in a relatively wide range of temperatures. The exception to this rule are just the isothermal entropy changes, which are induced by sufficiently large change of the magnetic field exceeding the saturation field [see the curve $\Delta h/J_1=6.0$ in Fig. \ref{fig4}(h)]. Contrary to this, the spin-1 Heisenberg diamond cluster with the interaction ratio $J_2/J_1=0.75$ shows an outstanding crossover between the inverse and conventional MCE. While the inverse MCE with $-\Delta S_m<0$ prevails at lower temperatures and magnetic-field changes, the conventional MCE with $-\Delta S_m>0$ dominates at higher temperatures and magnetic-field changes [see Fig. \ref{fig4}(e)-(f)].  

\begin{figure}
\begin{center}
\includegraphics[width=0.5\textwidth]{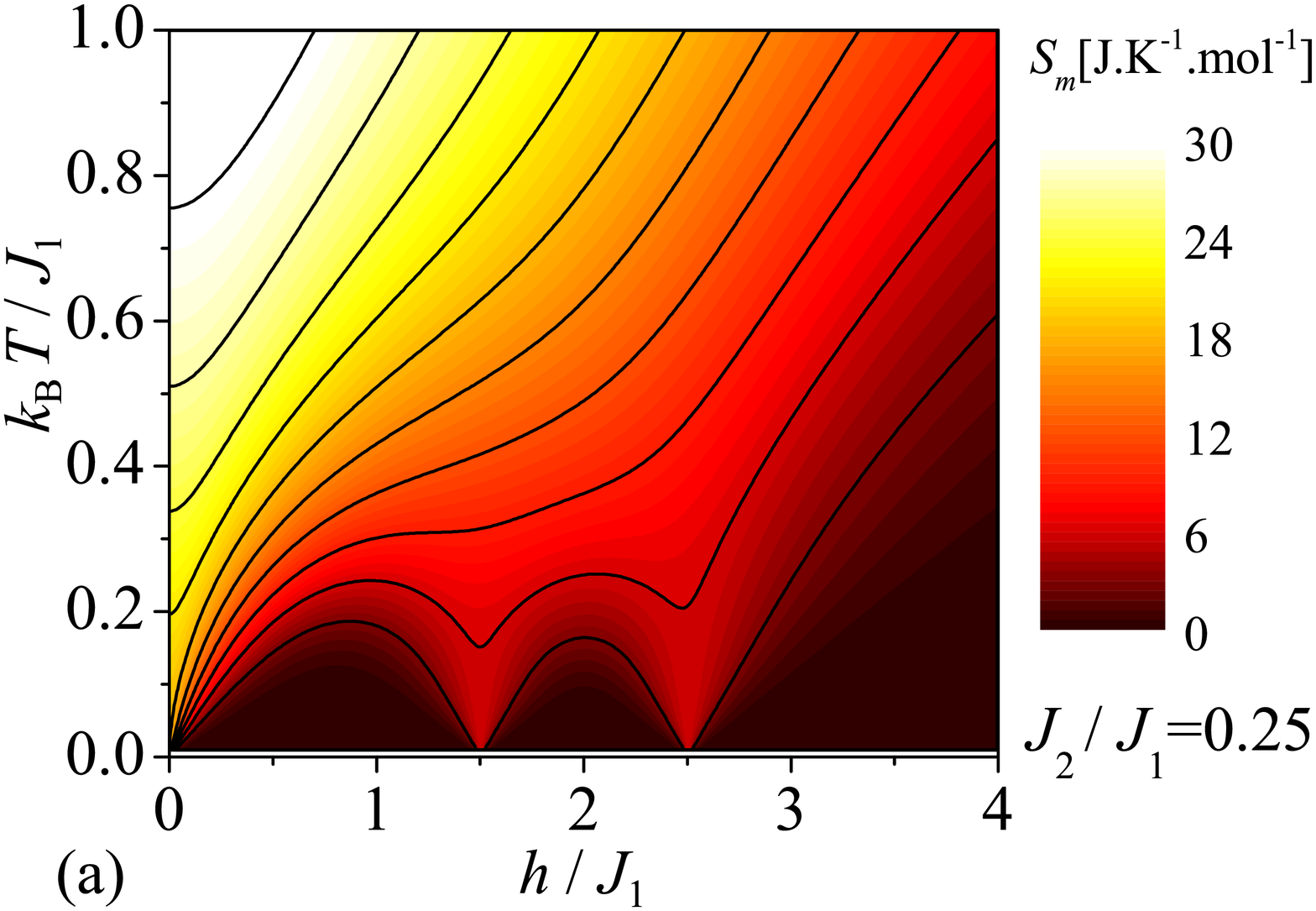}
\hspace{-0.7cm}
\includegraphics[width=0.5\textwidth]{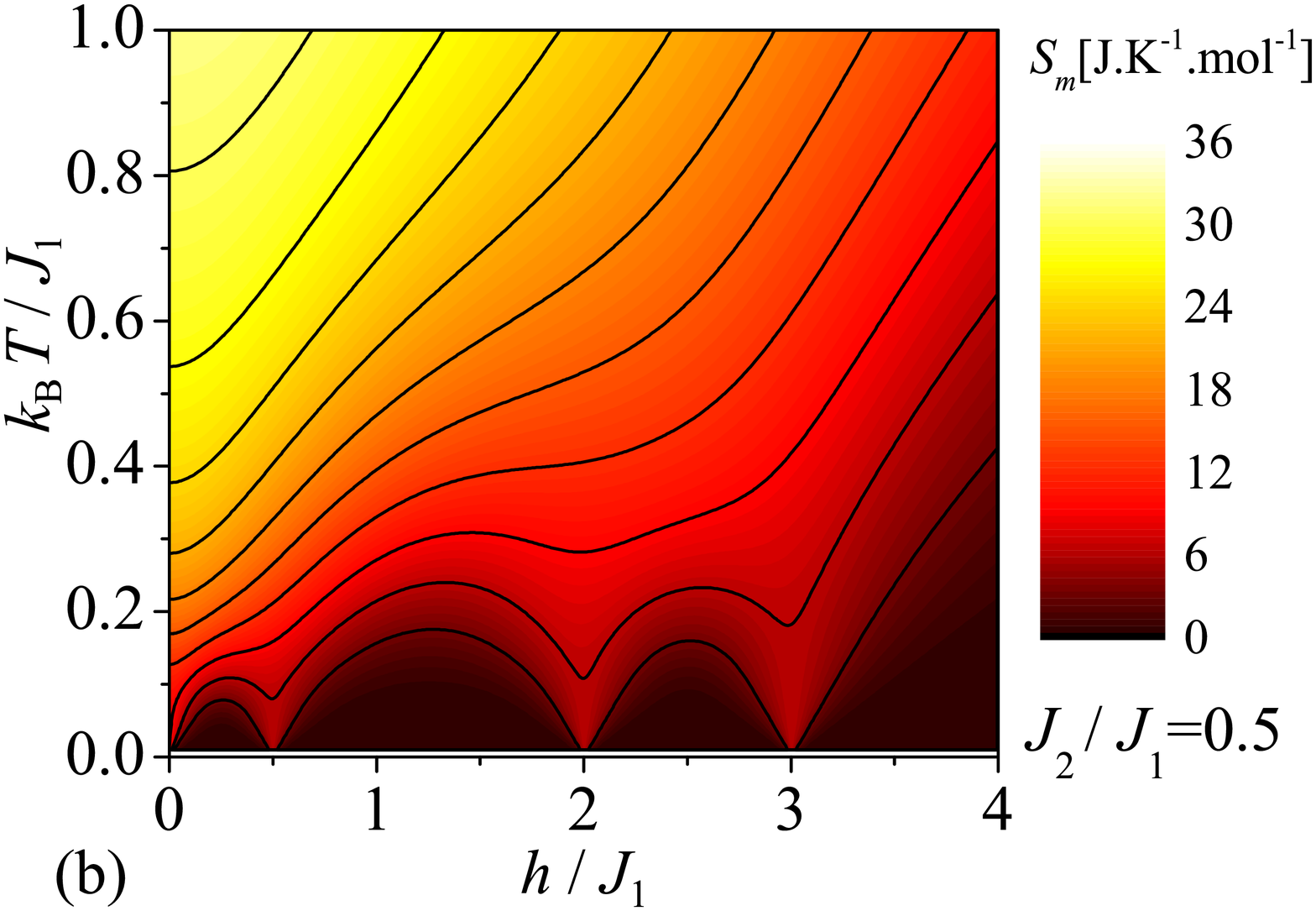}
\hspace{-0.7cm}
\includegraphics[width=0.5\textwidth]{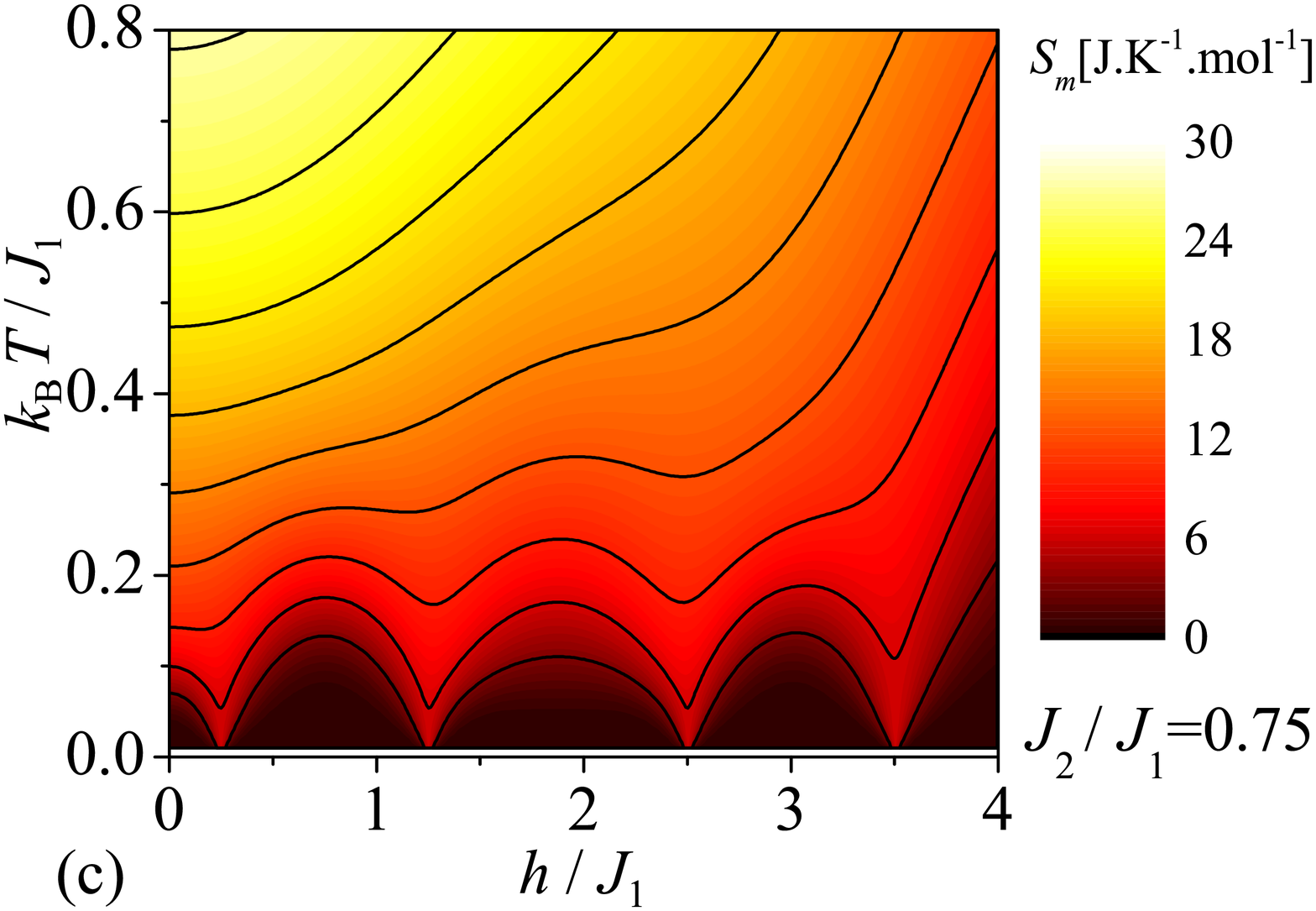}
\hspace{-0.2cm}
\includegraphics[width=0.5\textwidth]{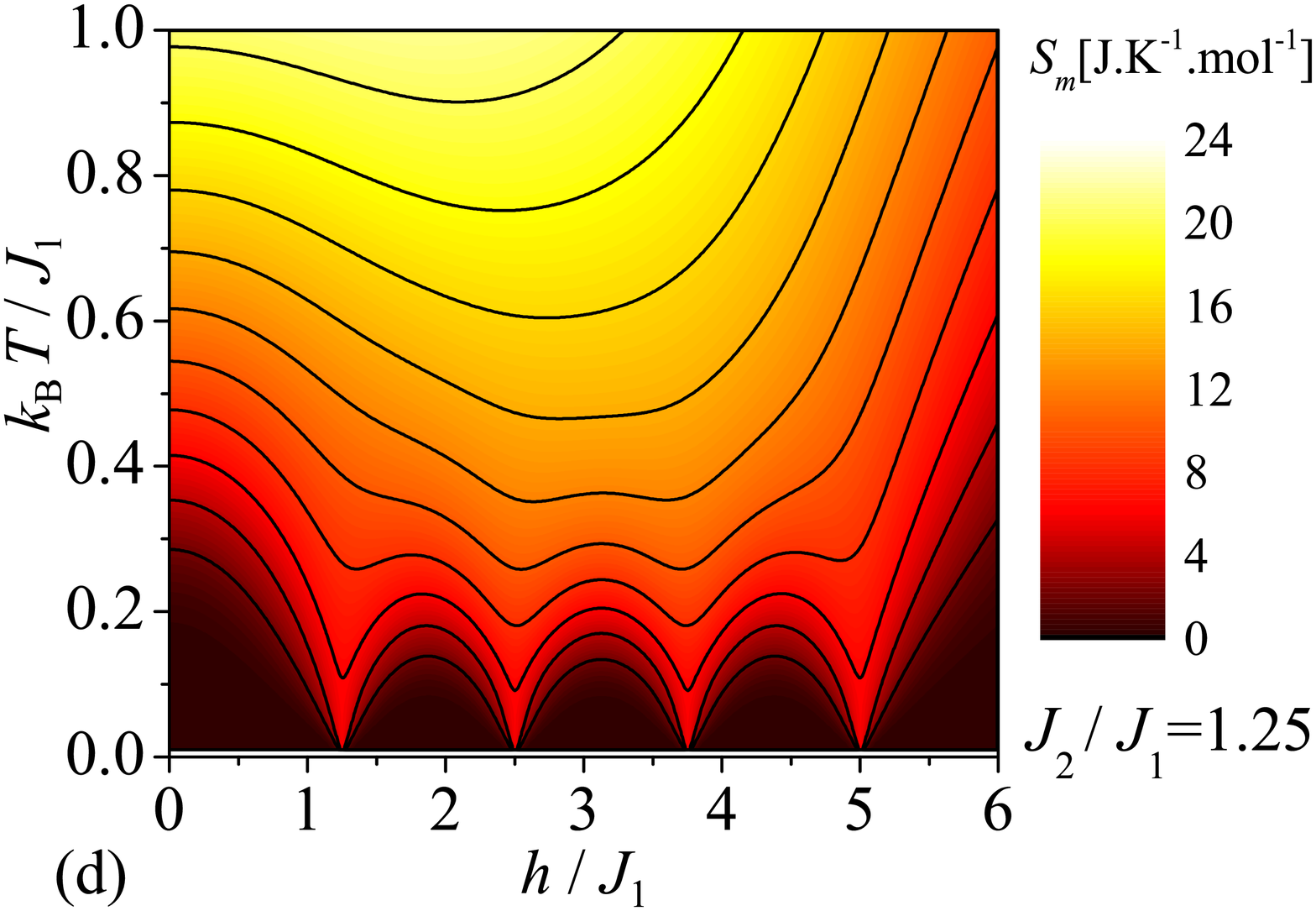}
\end{center}
\caption{A density plot of the molar entropy in J.K$^{-1}$.mol$^{-1}$ units in the magnetic field versus temperature plane for four different values of the interaction ratio:
(a) $J_2/J_1 = 0.25$; (b) $J_2/J_1 = 0.5$; (c) $J_2/J_1 = 0.75$; (d) $J_2/J_1 = 1.25$.}
\label{fig5}       
\end{figure}

Last but not least, let us examine the adiabatic change of temperature as another basic magnetocaloric property of the spin-1 Heisenberg diamond cluster. For this purpose, density plots of the molar entropy are displayed in Fig.~\ref{fig5}(a)-(d) in the magnetic field versus temperature plane for four selected values of the interaction ratio $J_2/J_1$, which have been  previously used in order to demonstrate a diversity of the magnetization profiles. It should be emphasized that black contour lines shown in Fig.~\ref{fig5}(a)-(d) correspond to isentropy lines, from which one can easily deduce adiabatic changes of temperature achieved upon lowering of the external magnetic field. It is quite evident from Fig.~\ref{fig5}(a)-(d) that the most notable changes of temperature occur in vicinity of all critical magnetic fields, whereas a sudden drop (rise) in temperature occurs during the adiabatic demagnetization slightly above (below) critical magnetic field. Hence, it follows that the abrupt magnetization jump manifest itself during the adiabatic demagnetization as a critical fan spread over a respective critical magnetic field. Two critical fans can be accordingly observed in Fig.~\ref{fig5}(a), three critical fans are visible in Fig.~\ref{fig5}(b) and four critical fans appear in Fig.~\ref{fig5}(c) and (d). It can be seen from Fig.~\ref{fig5}(a)-(d) that most of isentropes converge to some nonzero temperature as the external magnetic field gradually vanishes. More specifically, all isentropes of the spin-1 Heisenberg diamond cluster with the interaction ratio $J_2/J_1=0.75$ or 1.25 acquire nonzero temperature as the external magnetic field goes to zero [see Fig.~\ref{fig5}(c)-(d)]. This observation can be related with presence of zero-field singlet ground state $|0,2,2\rangle$, which is responsible for zero magnetization plateau. On the other hand, the spin-1 Heisenberg diamond cluster with the interaction ratio $J_2/J_1=0.25$ or 0.5 may exhibit during the adiabatic demagnetization a sizable drop of temperature down to ultra-low temperatures due to absence of zero magnetization plateau \cite{karl17}. To achieve this intriguing magnetocaloric feature, the molar entropy should be fixed to a smaller value than the entropy corresponding to a degeneracy of the respective zero-field ground state, i.e. $S_m < R\ln 9 \approx  18.3$ J.K$^{-1}$.mol$^{-1}$ for the zero-field ground state $|2,0,2\rangle$ emergent for $J_2/J_1=0.25$ or $S_m < R\ln 3 \approx 9.1$ J.K$^{-1}$.mol$^{-1}$ for the zero-field ground state $|1,1,2\rangle$ emergent for $J_2/J_1=0.5$, respectively. These findings could be of particular importance when the molecular compound \{Ni$_4$\} would be used for refrigeration at ultra-low temperatures.   
  
\section{Theoretical modeling of tetranuclear nickel complex \{Ni$_4$\}}
\label{experiment}

\begin{figure}
\begin{center}
\includegraphics[width=0.5\textwidth]{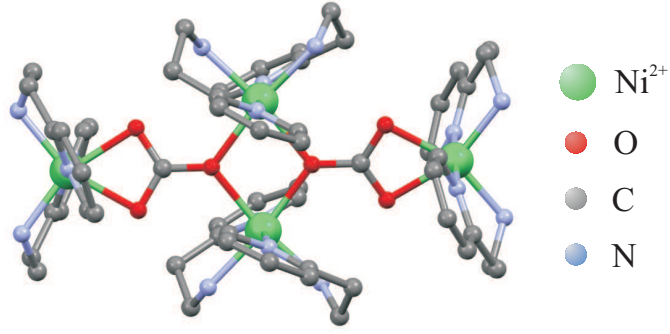}
\end{center}
\vspace{-0.5cm}
\caption{A crystal structure of the tetranuclear nickel complex [Ni$_4$($\mu$-CO$_3$)$_2$(aetpy)$_8$](ClO$_4$)$_4$ (aetpy = 2-aminoethyl-pyridine) abbreviated as \{Ni$_4$\} with a magnetic structure of 'butterfly tetramer', which is visualized according to crystallographic data reported in Ref.~\cite{escu98}. Crystallographic positions of hydrogen atoms and perchlorate anions ClO$_4^{-}$ were omitted for better clarity.}  
\label{figcs}
\end{figure}

\begin{figure}
\begin{center}
\includegraphics[width=0.50\textwidth]{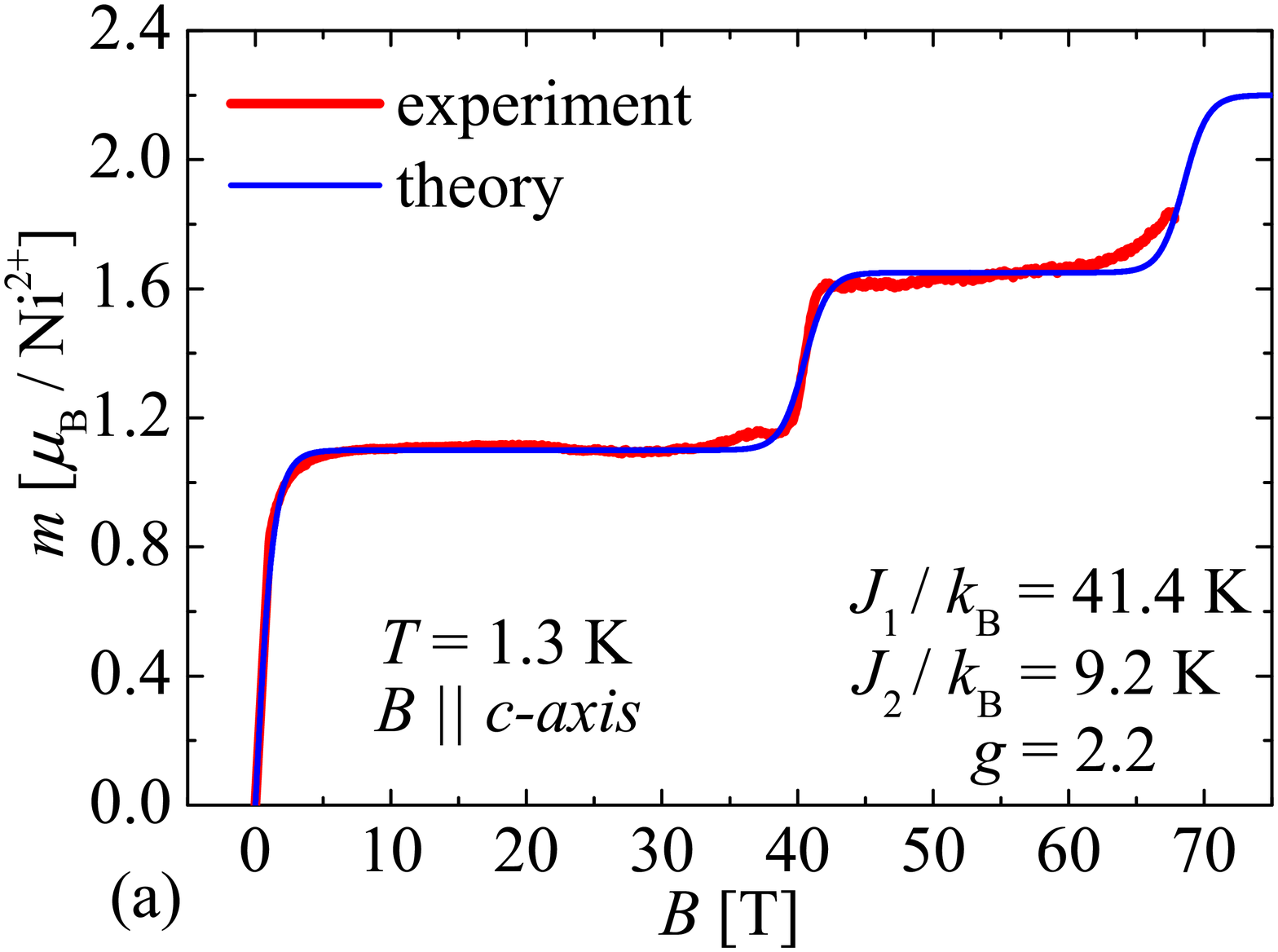}
\hspace{-0.8cm}
\includegraphics[width=0.50\textwidth]{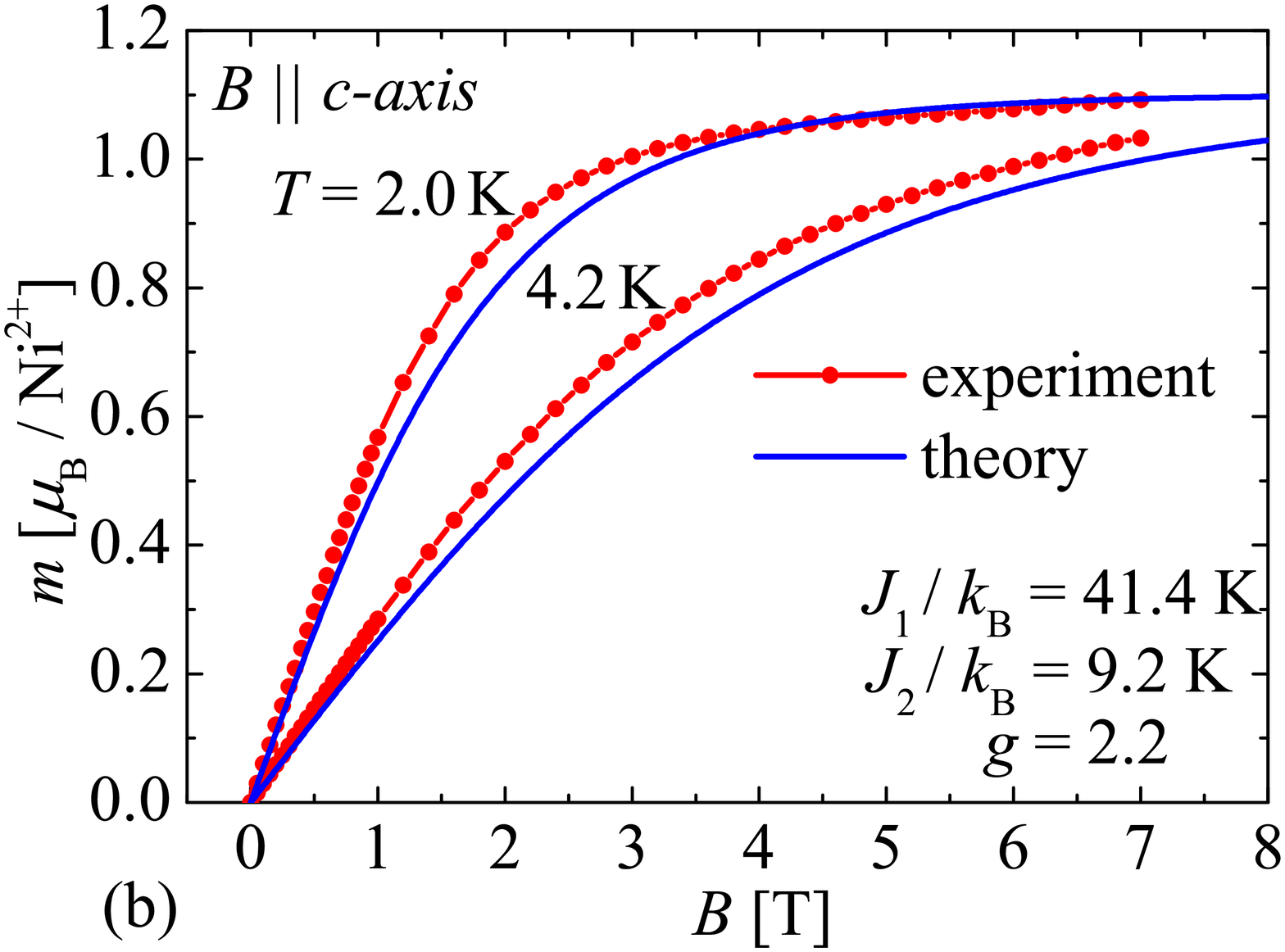}
\end{center}
\vspace{-0.5cm}
\caption{(a) High-field magnetization curve (red thick line) of the nickel complex \{Ni$_4$\} recorded in pulsed magnetic fields up to 68~T at the sufficiently low temperature 1.3~K and the respective theoretical fit (thin blue line) based on the spin-1 Heisenberg diamond cluster with the coupling constants $J_1/k_{\rm B} = 41.4$~K, $J_2/k_{\rm B} = 9.2$~K and the gyromagnetic factor $g=2.2$; (b) Magnetization curves (red lines with filled circles) of the nickel complex \{Ni$_4$\} in static magnetic fields up to 7~T at two different temperatures 2.0~K and 4.2~K versus the respective theoretical predictions (blue lines) for the spin-1 Heisenberg diamond cluster with the same set of the model parameters as specified in (a).}
\label{figmag}       
\end{figure}

In this part, we will interpret available experimental data for the magnetization and susceptibility of the tetranuclear nickel complex \{Ni$_4$\} \cite{escu98,hagi06}, which can be theoretically modeled by the spin-1 Heisenberg diamond cluster given by the Hamiltonian (\ref{hami}). It actually follows from Fig.~\ref{figcs} that the magnetic core of the tetranuclear coordination compound \{Ni$_4$\} constitutes a 'butterfly tetrameric' unit composed of four exchange-coupled Ni$^{2+}$ ions, which is formally identical with the magnetic structure of the spin-1 Heisenberg diamond cluster schematically illustrated in Fig.~\ref{fig1}. High-field magnetization data of the nickel complex \{Ni$_4$\} recorded in pulsed magnetic fields up to approximately 68~T at the sufficiently low temperature 1.3~K are presented in Fig.~\ref{figmag}(a) together with the respective theoretical fit based on the spin-1 Heisenberg diamond cluster. It is evident from Fig.~\ref{figmag}(a) that the measured magnetization data bear evidence of two wide intermediate plateaus roughly at 1.11 and 1.65 $\mu_{\rm B}$ per Ni$^{2+}$ ion, which are consistent with 1/2- and 3/4-plateaus when the total magnetization is scaled with respect to its saturation value and the appropriate value of the gyromagnetic factor $g=2.2$ of Ni$^{2+}$ ions is considered. The abrupt magnetization jumps detected at the critical magnetic fields $B_{c,1} \approx 40.5$~T and $B_{c,2} \approx 68.5$~T clearly delimit a width of these intermediate magnetization plateaus. The distinct magnetization profile with a sole presence of the intermediate 1/2- and 3/4-plateaus enables a simple estimation of the relevant coupling constants. First, it has been argued by the ground-state analysis that the intermediate 1/2- and 3/4-plateaus emerge in a zero-temperature magnetization curve as the only magnetization plateaus just if the interaction ratio falls into the range $J_2/J_1 \in (-1/2, 1/3)$. Second, one may take advantage of the fact that the width of 3/4-plateau $\Delta B_{3/4} = B_{c,2} - B_{c,1}$ is independent of the interaction ratio $J_2/J_1$ in contrast with the width of 1/2-plateau $\Delta B_{1/2} =  B_{c,1}$. Hence, the relative width of two magnetization plateaus $\delta_r = \Delta B_{3/4}:\Delta B_{1/2} = 28~{\rm T} : 40.5~{\rm T} \approx 0.69$ observed in experiment can be straightforwardly exploited for an unambiguous determination of a relative strength of the coupling constants:
\begin{eqnarray}
\delta_r = \frac{\Delta B_{3/4}}{\Delta B_{1/2}} = \frac{B_{c,2} - B_{c,1}}{B_{c,1}} = \frac{J_1}{2 J_2 + J_1} \quad \Longrightarrow \quad 
\frac{J_2}{J_1} = \frac{1 - \delta_r}{2 \delta_r} = \frac{2 B_{c,1} - B_{c,2}}{2(B_{c,2}-B_{c,1})} \approx \frac{2}{9}.
\label{eqdr}
\end{eqnarray}
Once determined, the absolute values of the coupling constants $J_1$ and $J_2$ can be easily calculated for instance from the first critical field $B_{c,1} = 40.5$~T when taking into account knowledge of the interaction ratio (\ref{eqdr}):
\begin{eqnarray}
g \mu_{\rm B} B_{c,1} = J_1 + 2 J_2 \approx \frac{13}{9} J_1 \quad \Longrightarrow \quad 
\frac{J_1}{k_{\rm B}} \approx  \frac{9 g \mu_{\rm B}}{13 k_{\rm B}} B_{c,1} = 41.4 \, {\rm K}, \quad
\frac{J_2}{k_{\rm B}} \approx  \frac{2 J_1}{9 k_{\rm B}} = 9.2 \, {\rm K}.   
\label{eqcc}
\end{eqnarray}
In accordance with this argumentation, the spin-1 Heisenberg diamond cluster with the coupling constants $J_1/k_{\rm B} = 41.4$~K, $J_2/k_{\rm B} = 9.2$~K and the gyromagnetic factor $g=2.2$ indeed satisfactorily reproduces the high-field magnetization data of the butterfly-tetramer compound \{Ni$_4$\} as convincingly evidenced by the respective theoretical fit, which is shown in Fig.~\ref{figmag}(a) as a thin blue line running through the experimental data. However, the magnetization curves of the tetranuclear nickel complex \{Ni$_4$\} measured in static magnetic fields up to 7~T at two different temperatures 2.0~K and 4.2~K are slightly underestimated by the respective theoretical fit based on the spin-1 Heisenberg diamond cluster with the assigned set (\ref{eqcc}) of the model parameters. It should be nevertheless mentioned that any change of the coupling constants $J_1$ and $J_2$ does not significantly improve a theoretical fit of these experimental data. It has been found in Ref. \cite{hagi06} that the significant improvement of the theoretical fit can be achieved only when considering a weak ferromagnetic exchange coupling $J_3/k_{\rm B} = -0.66$~K between the further-distant spins $S_3$ and $S_4$, which allows a steeper uprise of the magnetization in a low-field range. A consideration of the exchange coupling between the further-distant spins $S_3$ and $S_4$ is however beyond the scope of the present article.         

\begin{figure}
\begin{center}
\includegraphics[width=0.50\textwidth]{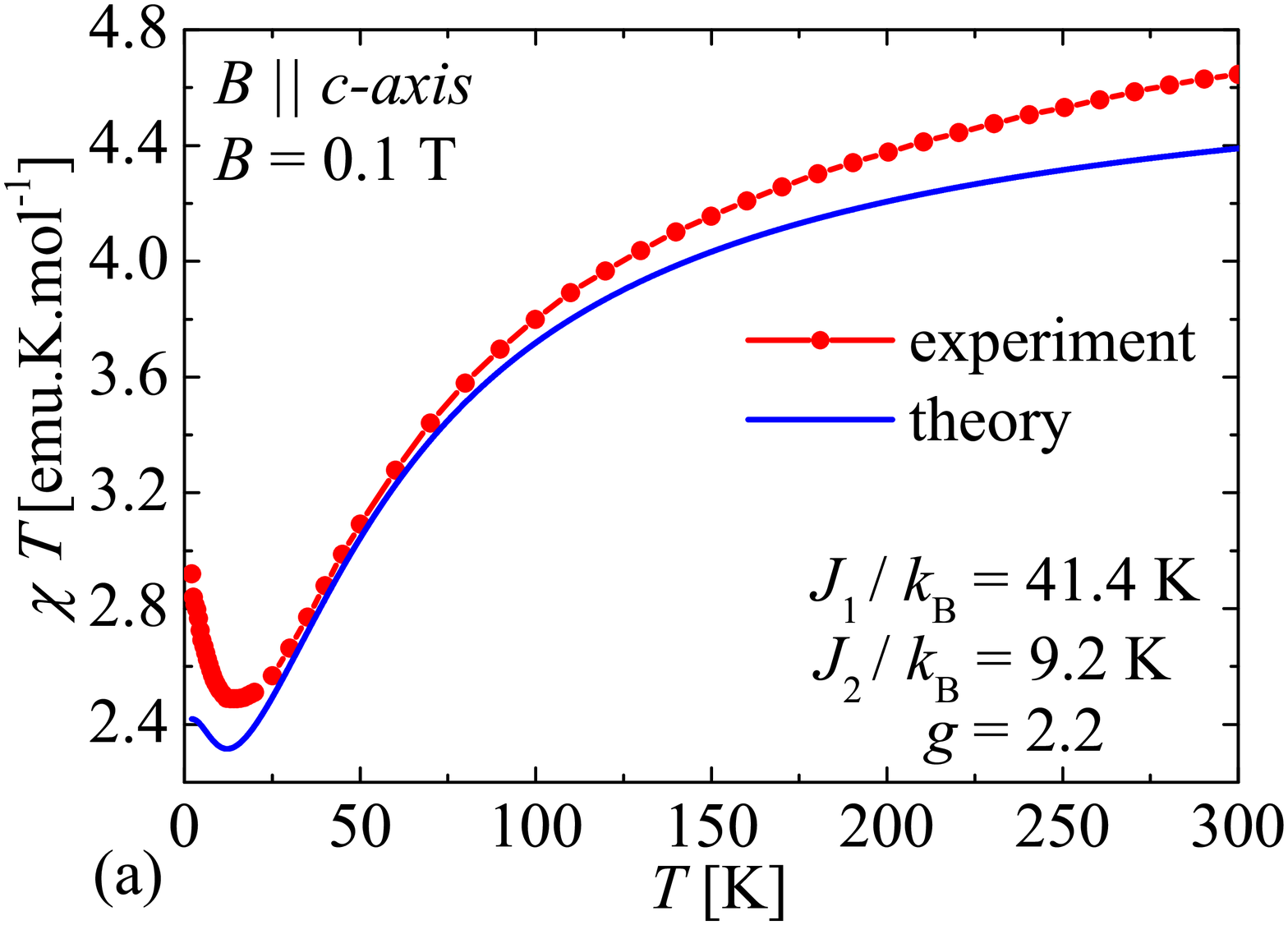}
\hspace{-0.5cm}
\includegraphics[width=0.50\textwidth]{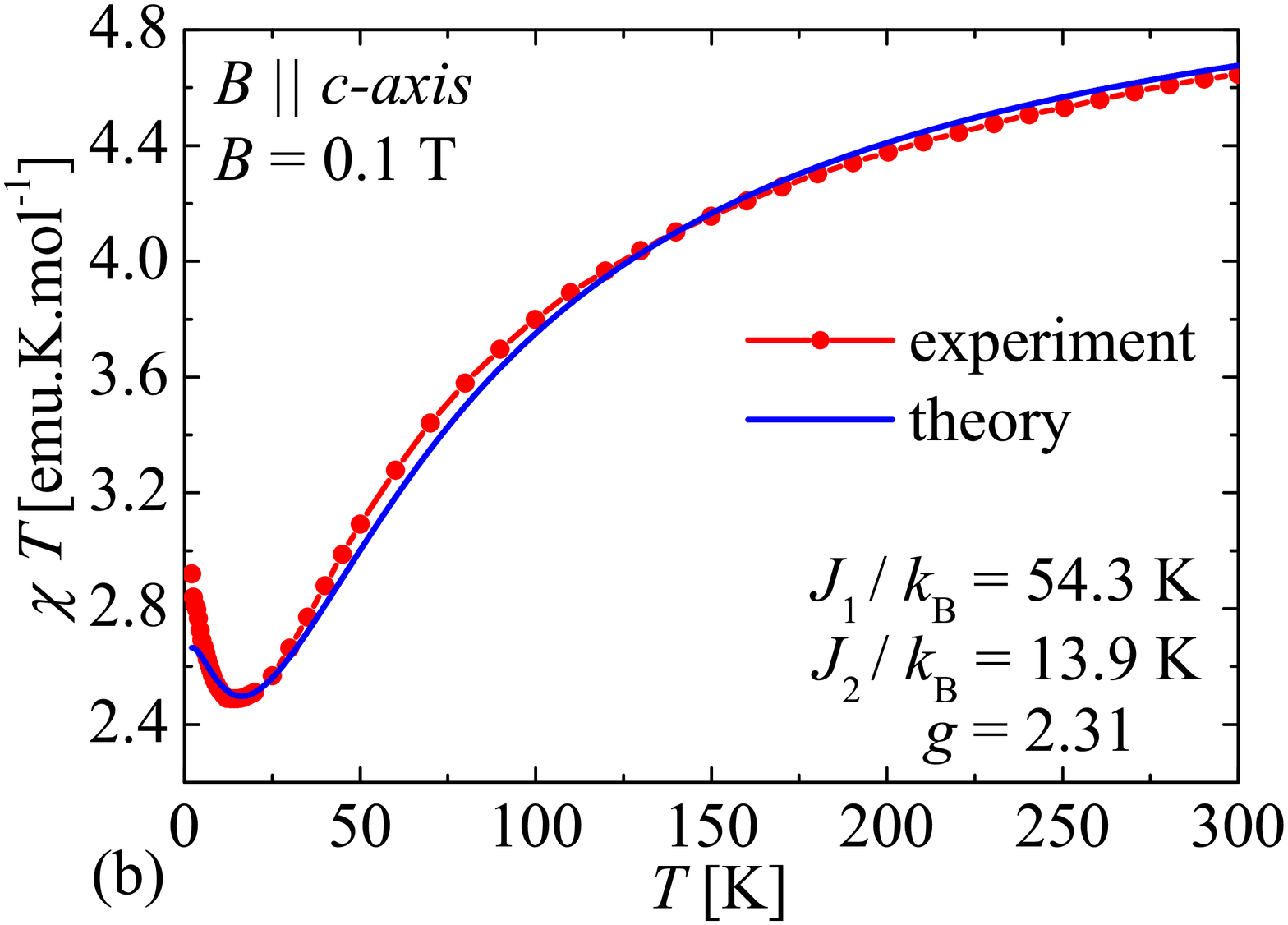}
\end{center}
\vspace{-0.5cm}
\caption{Temperature variations of the susceptibility times temperature product of the nickel complex \{Ni$_4$\} (a red line with filled circles) at the magnetic field $B = 0.1$~T versus the respective theoretical prediction based on the spin-1 Heisenberg diamond cluster (a blue line) when considering two different sets of the fitting parameters. The fitting set specified in the panel (a) is taken without any further adjustment from the aforedescribed fitting procedure of the high-field magnetization data, whereas the fitting set specified in the panel (b) represents the best fit of the susceptibility data.}
\label{figsus}       
\end{figure}

Next, we will employ the coupling constants (\ref{eqcc}) ascribed to the coordination compound \{Ni$_4$\} for a theoretical interpretation of a temperature dependence of the susceptibility times temperature ($\chi T$) product. To this end, the available experimental data for the $\chi T$ product of the tetranuclear nickel complex \{Ni$_4$\} are confronted in Fig.~\ref{figsus}(a) with the respective theoretical prediction based on the spin-1 Heisenberg diamond cluster by assuming the model parameters (\ref{eqcc}) previously extracted from the fitting procedure of the high-field magnetization data. Although a theoretical curve qualitatively captures all essential features for temperature variations of the $\chi T$ product including a local minimum experimentally observed around 14~K, the good quantitative accordance between the experimental and theoretical data is found just in a relatively narrow range of temperatures $T \in (25, 80)$~K while outside of this temperature range the theoretical data generally underestimate the experimental ones. We have therefore adapted the optimization technique based on a hill-climbing procedure in order to find the best fitting set for the $\chi T$ data. This procedure provided for the tetranuclear nickel compound \{Ni$_4$\} described by the spin-1 Heisenberg diamond cluster another fitting set of the model parameters $J_1/k_{\rm B} = 54.3$~K, $J_2/k_{\rm B} = 13.9$~K and $g=2.31$, which not only qualitatively but also quantitatively captures the experimental data in a full range of temperatures as exemplified in Fig.~\ref{figsus}(b). While the rise of the gyromagnetic factor by a few percent (cca. 5 \%) could be attributed to a substantial temperature difference within the magnetization and susceptibility measurements, the relatively large discrepancy in assessment of both coupling constants clearly indicates an oversimplified nature of the spin-1 Heisenberg diamond-cluster model given by the Hamiltonian (\ref{hami}). It is quite reasonable to conjecture from nearly isotropic character of the magnetization curves measured along two orthogonal crystallographic axes \cite{hagi06} that the axial and/or rhombic zero-field-splitting parameters acting on Ni$^{2+}$ ions are presumably negligible and hence, the discrepancies in the magnetization and susceptibility data could be resolved when taking into consideration the biquadratic interaction and/or the pair exchange interaction between the further-distant spins $S_3$ and $S_4$.        

\begin{figure}
\begin{center}
\includegraphics[width=0.50\textwidth]{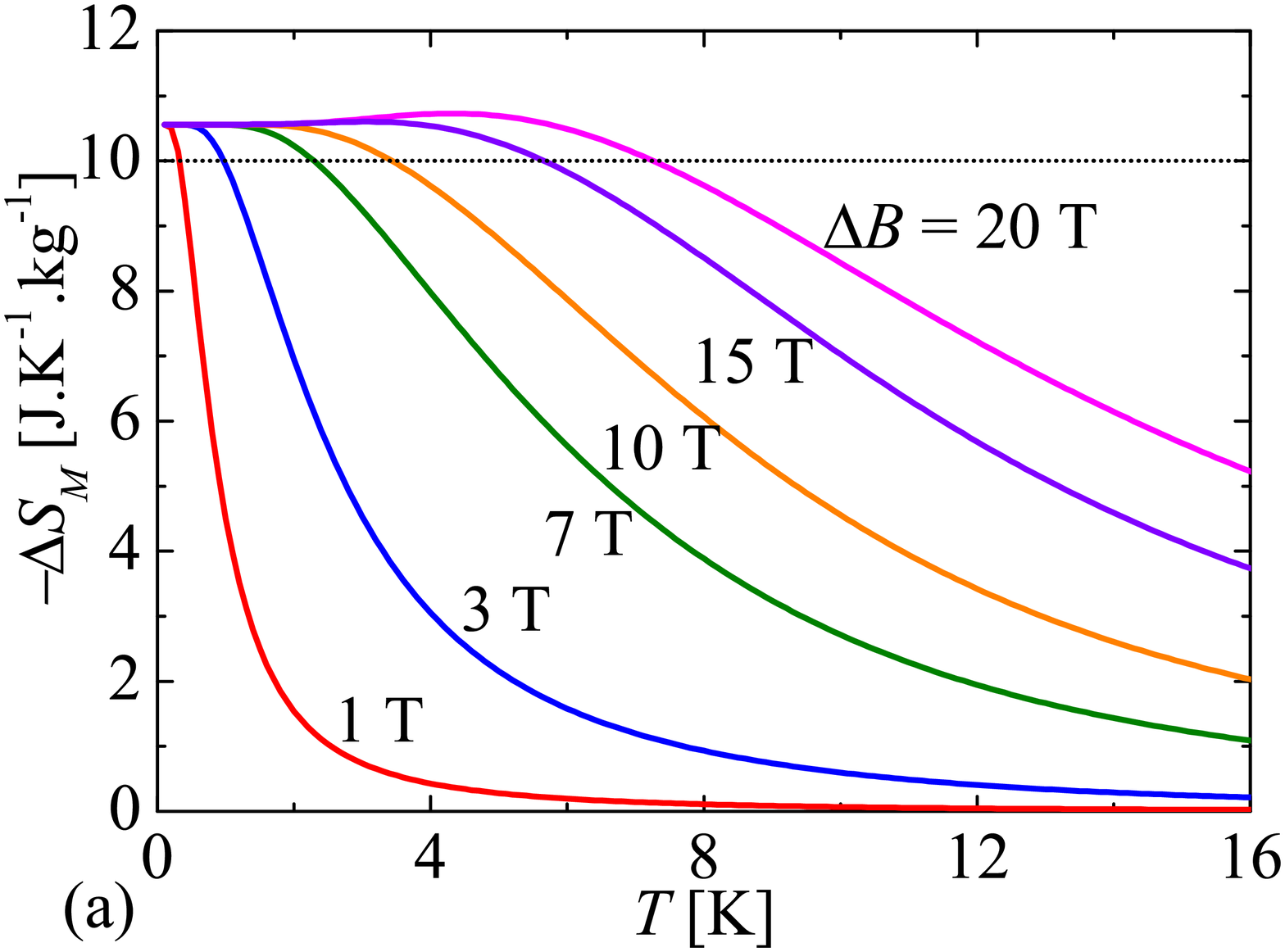}
\hspace{-0.5cm}
\includegraphics[width=0.50\textwidth]{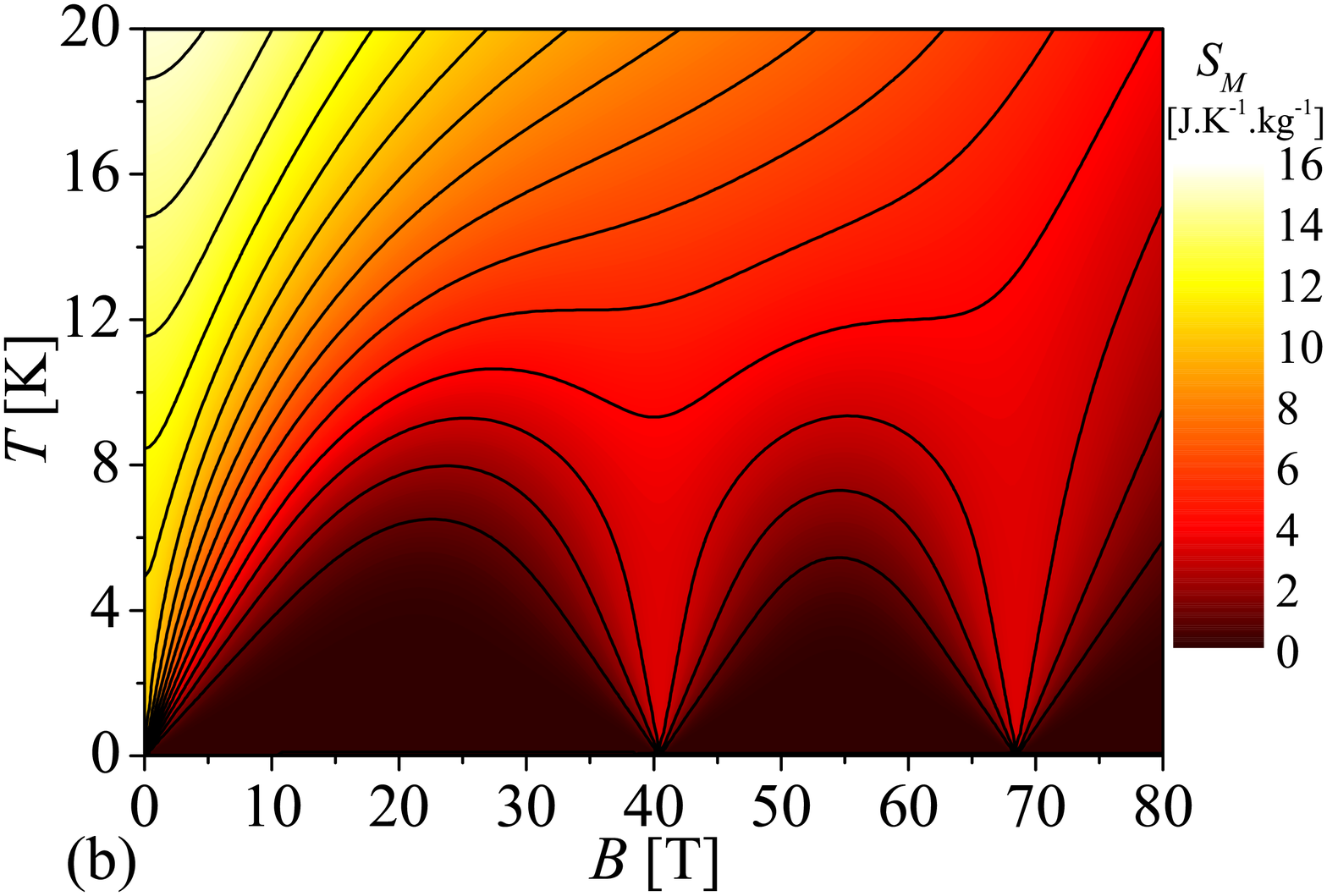}
\end{center}
\vspace{-0.5cm}
\caption{(a) Temperature variations of the isothermal mass entropy change $-\Delta S_M$ of the tetranuclear nickel complex \{Ni$_4$\} for a few selected values of the magnetic-field change. A dotted line, which serves as a guide for eyes only, delimits the parameter space with the enhanced MCE $-\Delta S_M > 10$ J.K$^{-1}$.kg$^{-1}$; (b) A density plot of the entropy in the field-temperature plane. Displayed contour lines illustrate adiabatic changes of temperature achieved upon variation of the magnetic field. All results presented in Fig. \ref{figent}(a) and (b)  were calculated for the spin-1 Heisenberg diamond cluster with the coupling constants $J_1/k_{\rm B} = 41.4$~K, $J_2/k_{\rm B} = 9.2$~K and the gyromagnetic factor $g=2.2$.}
\label{figent}       
\end{figure}

Last but not least, the best fitting set (\ref{eqcc}) extracted for the spin-1 Heisenberg diamond-cluster model from the high-field magnetization curve of the tetranuclear nickel complex \{Ni$_4$\} will be used for making a theoretical prediction of its basic magnetocaloric properties not reported experimentally hitherto. More specifically, we will investigate in detail temperature variations of the isothermal magnetic entropy change as well as field-induced changes of temperature during the adiabatic demagnetization. It is evident from Fig. \ref{figent}(a) that the isothermal mass entropy change of the nickel compound \{Ni$_4$\} gradually diminishes from its maximum value $-\Delta S_M \approx 10.6$ J.K$^{-1}$.kg$^{-1}$ upon increasing of temperature whenever the magnetic-field change is sufficiently small $\Delta B < 15$~T. On assumption that the magnetic-field change is set $\Delta B = 7$~T the molecular compound \{Ni$_4$\} provides an efficient refrigerant below 2.3~K with the enhanced MCE $-\Delta S_M > 10$ J.K$^{-1}$.kg$^{-1}$. It should be stressed that a subtle rise of the isothermal entropy change $-\Delta S_M$ can be detected for the higher magnetic-field changes [e.g. see the curve for $\Delta B = 20$~T in Fig. \ref{figent}(a)], which is however of very limited applicability for the cooling technologies. 

On the other hand, the density plot of the magnetic mass entropy in the magnetic field versus temperature plane is displayed in Fig. \ref{figent}(b) with the aim to elucidate a parameter space suitable for cooling purposes. The relevant contour lines with constant magnetic entropy bring insight into magnetic-field driven changes of temperature during the process of adiabatic demagnetization. A considerable drop and rise of temperature apparently occurs in the isentropes near the critical magnetic fields, which correspond to the magnetic-field-driven magnetization jumps. If the magnetic entropy is set sufficiently close to the particular value $S_M \approx 10.6$ J.K$^{-1}$.kg$^{-1}$, moreover,  the adiabatic demagnetization should cause a sizable drop of temperature of the molecular complex \{Ni$_4$\} with up to $-\Delta T \approx 10$~K achieved due to the magnetic-field change $\Delta B = 7$~T.      

\section{Conclusions}
\label{conclusion}

In the present article we have investigated in detail magnetic and magnetocaloric properties of the spin-1 Heisenberg diamond cluster with two different coupling constants through an exact diagonalization based on the Kambe's method, which takes advantage of a local conservation of composite spins formed by spin-1 entities located in opposite corners of a diamond spin cluster. It has been verified that the spin-1 Heisenberg diamond cluster exhibits several intriguing quantum ground states, which come to light in low-temperature magnetization curves as intermediate 1/4-, 1/2- or 3/4-plateau depending on a specific choice of the interaction ratio and the magnetic field. We have demonstrated a substantial diversity of the magnetization curves, which may exhibit different magnetization profiles with either a single 3/4-plateau, a sequence of two consecutive 1/2- and 3/4-plateaus, three consecutive 1/4-, 1/2- and 3/4-plateaus, four consecutive 0-, 1/4-, 1/2- and 3/4-plateaus or is completely free of any plateau. In addition, the spin-1 Heisenberg diamond cluster may also exhibit the enhanced MCE, which may be relevant for a low-temperature refrigeration achieved through the adiabatic demagnetization on assumption that a relative strength of the coupling constants $J_2/J_1 \in (-1, 2/3)$ is consistent with absence of the zero magnetization plateau. 

It has been evidenced that the spin-1 Heisenberg diamond cluster with the antiferromagnetic coupling constants $J_1$/$k_{\rm B}$ =  41.4~K, $J_2$/$k_{\rm B}$ = 9.2~K and the gyromagnetic factor $g=2.2$ satisfactorily captures low-temperature magnetization curves recorded for the tetranuclear nickel complex \{Ni$_4$\} including a size and position of the intermediate 1/2- and 3/4-plateaus \cite{hagi06}. Moreover, it turns out that the fractional magnetization plateaus observed experimentally bear evidence of two remarkable valence-bond-crystal ground states with either a single or double valence bond between the near-distant spin-1 Ni$^{2+}$ ions. It has been also suggested that the molecular compound \{Ni$_4$\} may provide a prospective cryogenic coolant with the maximal isothermal entropy change $-\Delta S_M = 10.6$ J.K$^{-1}$.kg$^{-1}$ suitable for a low-temperature refrigeration below 2.3~K.

\begin{acknowledgments}
This research was funded by The Ministry of Education, Science, Research and Sport of the Slovak Republic under the grant number VEGA 1/0105/20 and by The Slovak Research and Development Agency under the grant number APVV-18-0197. The authors thank Yasuo Narumi and Koichi Kindo for measuring high-field magnetization of the title compound.
\end{acknowledgments}


\begin{thebibliography}{999}
\bibitem{carl86} 
Carlin, R.L. 
{\em Magnetochemistry}; 
Springer: Berlin, Germany, 1986.

\bibitem{kahn93} 
Kahn, O. 
{\em Molecular Magnetism}; 
Wiley–VCH: New York, USA, 1993.

\bibitem{mill02}
Miller, J. S., Drillon, M. (Eds.). 
{\em Magnetism: Molecules to Materials I-V}; 
Wiley-VCH: Weinheim, Germany, 2002.

\bibitem{siek17} 
Sieklucka, B., Pinkowicz, D. (Eds.).
{\em Molecular Magnetic Materials}; 
Wiley–VCH: Weinheim, Germany, 2017.
								
\bibitem{gatt03} 
Gatteschi, D., Sessoli, R., 
Quantum Tunneling of Magnetization and Related Phenomena in Molecular Materials. 
{\em Angew. Chem. Int. Ed.} {\bf 2003}, {\em 42}, 268--297.

\bibitem{roch05} 
Rocha, A. R., Garc\'ia-su\'arez, V. M., Bailey, S. W., Lambert, C. J., Ferrer J., Sanvito, S. 
Towards molecular spintronics. 
{\em Nature Mater.} {\bf 2005}, {\em 4}, 335--339.

\bibitem{boga08} 
Bogani, L., Wernsdorfer, W. 
Molecular spintronics using single-molecule magnets. 
{\em Nature Mater.} {\bf 2008}, {\em 7}, 179--186.

\bibitem{urda11} 
Urdampilleta, M., Klyatskaya, S., Cleuziou, J.-P., Ruben, M., Wernsdorfer, W. 
Supramolecular spin valves. 
{\em Nature Mater.} {\bf 2011}, {\em 10}, 502--506.

\bibitem{leue01} 
Leuenberger, M.N., Loss D. 
Quantum computing in molecular magnets. 
{\em Nature} {\bf 2001}, {\em 410}, 789--793.

\bibitem{teja01} 
Tejada, J., Chudnovsky, E. M., Del Barco, E., Hernandez, J. M., Spiller, T. P. 
Magnetic qubits as hardware for quantum computers. 
{\em Nanotechnology} {\bf 2001}, {\em 12}, 181.

\bibitem{timc11}
Timco, G. A., Faust, T. B., Tuna, F., Winpenny, R. E. P. 
Linking heterometallic rings for quantum information processing and amusement. 
{\em Chem. Soc. Rev.} {\bf 2011}, {\em 40}, 3067--3075.

\bibitem{troi11}
Troiani, F., Affronte, M. 
Molecular spins for quantum information technologies.
{\em Chem. Soc. Rev.} {\bf 2011}, {\em 40}, 3119--3129.

\bibitem{esca18} 
Escalera-Moreno, L.,  Baldov, J. J., Gaita-Ario, A., Coronado, E.
Spin states, vibrations and spin relaxation in molecular nanomagnets and spin qubits: a critical perspective.
{\em Chem. Sci.} {\bf 2018}, {\em 9}, 3265--3275.

\bibitem{gait19} 
Gaita-Ario, A., Luis, F., Hill, S., Coronado, E.
Molecular spins for quantum computation.
{\em Nat. Chem.} {\bf 2019}, {\em 11}, 301--309.

\bibitem{guwu20} 
Gu, L., Wu, R. 
Origins of Slow Magnetic Relaxation in Single-Molecule Magnets.
{\em Phys. Rev. Lett.} {\bf 2020}, {\em 125}, 117203.

\bibitem{kind10}
Kindo, K., Takeyama, S., Tokunaga, M., Matsuda, Y. H., Kojima, E., Matsuo, A., Kawaguchi, K., Sawabe, H.
The International MegaGauss Laboratory at ISSP, The University of Tokyo. 
{\em J. Low Temp. Phys.} {\bf 2010}, {\em 159}, 381--388.

\bibitem{lipe12}
Li, L., Peng, T., Xiao, H. X., Lv, Y. L., Pan, Y., Herlach, F.
Magnet development program at the WHMFC.
{\em IEEE Trans. Appl. Supercond.} {\bf 2012}, {\em 22}, 4300304.

\bibitem{zher13}
Zherlitsyn, S., Wustmann, B., Herrmannsdorfer, T., Wosnitza, J.
Magnet-technology development at the dresden high magnetic field laboratory. 
{\em J. Low Temp. Phys.} {\bf 2013}, {\em 170}, 447--451.

\bibitem{hagi13}
Hagiwara, M., Kida, T., Taniguchi, K., Kindo, K.
Present Status and Future Plan at High Magnetic Field Laboratory in Osaka University.
{\em J. Low Temp. Phys.} {\bf 2013}, {\em 170}, 531--540.

\bibitem{maed14}
Maeda, H., Yanagisawa, Y.
Recent Developments in High-Temperature Superconducting Magnet Technology (Review).
{\em IEEE Trans. Appl. Supercond.} {\bf 2014}, {\em 24}, 4602412.

\bibitem{bear18}
 B\'eard, J., Billette, J., Ferreira, N., Frings, P., Lagarrigue, J.-M., Lecouturier, F., Nicolin, J.-P.
Design and Tests of the 100-T Triple Coil at LNCMI. 
{\em IEEE Trans. Appl. Supercond.} {\bf 2018}, {\em 28}, 4300305.

\bibitem{hahn19}
Hahn, S., Kim, K., Kim, K., Hu, X., Painter, T., Dixon, T., Kim, S., Bhattarai, K. R., Noguchi, S., Jaroszynski, J., Larbalestier, D. C. 
45.5--tesla direct-current magnetic field generated with a high-temperature superconducting magnet.
{\em Nature} {\bf 2019}, {\em 570}, 496--499.

\bibitem{mich20}
Michel, J. R., Nguyen, D. N., Lucero, J. D. 
Design, Construction, and Operation of New Duplex Magnet at Pulsed Field Facility-NHMFL.
{\em IEEE Trans. Appl. Supercond.} {\bf 2020}, {\em 30}, 8977361.

\bibitem{taki11}
Takigawa, M., Mila, F. 
Magnetization Plateaus. In {\em Introduction to Frustrated Magnetism}; 
Lacroix, C., Mendels, Ph., Mila, F., Eds.; Springer-Verlag: Berlin, Germany, 2011; pp. 241--267.

\bibitem{hida05} 
Hida, K., Affleck, I.
Quantum vs Classical Magnetization Plateaus of S = 1/2 Frustrated Heisenberg Chains.
{\em J. Phys. Soc. Jpn.} {\bf 2005}, {\em 74}, 1849. 

\bibitem{cole13}
Coletta, T., Zhitomirsky, M.E., Mila, F.
Quantum stabilization of classically unstable plateau structures.
{\em Phys. Rev. B} {\bf 2013}, {\em 87}, 060407.

\bibitem{karl17} 
Karl'ov\'a, K., Stre\v{c}ka, J., Richter, J.
Enhanced magnetocaloric effect in the proximity of magnetization steps and jumps of spin-1/2 XXZ Heisenberg regular polyhedra.
{\em J. Phys.: Condens. Matter} {\bf 2017}, {\em 29}, 125802.

\bibitem{karl20} 
Karl'ov\'a, K.
Spin-1/2 XXZ Heisenberg cupolae: magnetization process and related enhanced magnetocaloric effect.
{\em Condens. Matter Phys.} {\bf 2020}, {\em 23}, in press.

\bibitem{naru98}
Narumi, Y., Sato, R., Kindo, K., Hagiwara, M.
Magnetic property of an S = 1 antiferromagnetic dimer compound.
{\em J. Magn. Magn. Mater.} {\bf 1998}, {\em 177-181}, 685--686.

\bibitem{stre05}
Stre\v{c}ka, J., Ja\v{s}\v{c}ur, M., Hagiwara, M., Narumi, Y., Kuch\'ar, J., Kimura, S., Kindo K. 
Magnetic behavior of a spin-1 dimer: model system for homodinuclear nickel(II) complexes.
{\em J. Phys. Chem. Solids} {\bf 2005}, {\em 66}, 1828--1837.

\bibitem{stre08}
Stre\v{c}ka, J., Hagiwara, M., Bal\'a\v{z}, P., Ja\v{s}\v{c}ur, M., Narumi, Y., Kimura, S., Kuch\'ar, J., Kindo, K. 
Breakdown of an intermediate plateau in the magnetization process of anisotropic spin-1 Heisenberg dimer: Theory vs. experiment.
{\em Physica B} {\bf 2008}, {\em 403}, 3146--3153.

\bibitem{hagi99}
Hagiwara, M., Narumi, Y., Minami, K., Tatani, K., Kindo, K.
Magnetization Process of the S = 1/2 and 1 Ferrimagnetic Chain and Dimer.
{\em J. Phys. Soc. Jpn.} {\bf 1999}, {\em 68}, 2214--2217.

\bibitem{choi08}
Choi, K.-Y., Dalal, N. S., Reyes, A. P., Kuhns, Ph. R., Matsuda, Y. H., Nojiri, H., Mal, S. S., Kortz, U.
Pulsed-field magnetization, electron spin resonance, and nuclear spin-lattice relaxation in the \{Cu$_3$\} spin triangle.
{\em Phys. Rev. B} {\bf 2008}, {\em 77}, 024406. 

\bibitem{pono15}
Ponomaryov, A. N., Kim, N., Jang, Z. H., van Tol, J., Koo, H. J., Law, J. M., Suh, B. J., Yoon, S., Choi, K. Y.
Spin decoherence processes in the S = 1/2 scalene triangular cluster (Cu$_3$(OH)).
{\em New J. Phys.} {\bf 2015}, {\em 17}, 033042.

\bibitem{chat19}
Chattopadhyay, S. Lenz, B., Kanungo, S., Sushila, Panda, S. K., Biermann, S., Schnelle, W., Manna, K., Kataria, R., Uhlarz, M., Skourski, Y., Zvyagin, S. A., Ponomaryov, A., 
Herrmannsd\"o rfer, H., Patra, R., Wosnitza, J.
Pronounced 2/3 magnetization plateau in a frustrated S = 1 isolated spin-triangle compound: Interplay between Heisenberg and biquadratic exchange interactions.
{\em Phys. Rev. B} {\bf 2019}, {\em 100}, 094427.

\bibitem{mull00}
M\"uller, A., Beugholt, C.,  K\"ogerler, P., Bolgge, H., Bud'ko, S., Luban, M. 
[Mo$_{12}$O$_30$($\mu$-OH)$_{10}$H$_2$\{Ni(H$_2$O)$_3$\}$_4$], a highly symmetrical $\epsilon$-Keggin unit capped with four Ni(II) centers: Synthesis and magnetism.
{\em Inorg. Chem.} {\bf 2000}, {\em 39}, 5176--5177.

\bibitem{post05}
Postnikov, A. V., Br\"uger, M., Schnack, J.
Exchange interactions and magnetic anisotropy in the 'Ni$_4$' magnetic molecule.
{\em Phase Transit.} {\bf 2005}, {\em 78}, 47--59.

\bibitem{schn06}
Schnack, J., Br\"uger, M., Luban, M., K\"ogerler, P., Morosan, E., Fuchs, R., Modler, R., Nojiri, H., Rai, R. C., Cao, J., Musfeldt, J. L., Wei, X.,
Observation of field-dependent magnetic parameters in the magnetic molecule \{Ni$_4$Mo$_{12}$\}.
{\em Phys. Rev. B} {\bf 2006}, {\em 73}, 094401.

\bibitem{zbli08}
Li, Z. B., Yao, K. L., Liu, Z. L. 
Thermodynamic properties of a spin-1 tetrahedron as a model for a molecule-based compound [Mo$_{12}$O$_30$($\mu$-OH)$_{10}$H$_2$\{Ni(H$_2$O)$_3$\}$_4$] $\cdot$ 14 H$_2$O.
{\em J. Magn. Magn. Mater.} {\bf 2008}, {\em 320}, 1759--1764.

\bibitem{nath13}
Nath, R., Tsirlin, A. A., Khuntia, P., Janson, O., F\"orster, T., Padmanabhan, M., Li, J., Skourski, Yu., Baenitz, M., Rosner, H., Rousochatzakis, I.
Magnetization and spin dynamics of the spin S=1/2 hourglass nanomagnet Cu$_5$(OH)$_2$(NIPA)$_4$ $\cdot$ 10H$_2$O.
{\em Phys. Rev. B} {\bf 2013} {\em 87}, 214417.

\bibitem{szal20}
Sza\l{}owski, K., Kowalewska, P.
Magnetocaloric Effect in Cu5-NIPA Molecular Magnet: A Theoretical Study.
{\em Materials} {\bf 2020}, {\em 13}, 485.

\bibitem{luba02}
Luban, M., Borsa, F., Bud'ko, S., Canfield, P., Jun, S., Jung, J. K., K\"ogerler, P., Mentrup, D., M\"uller, A., Modler, R., Procissi, D., Suh, B. J., Torikachvili, M.
Heisenberg spin triangles in \{V$_6$\}-type magnetic molecules: Experiment and theory.
{\em Phys. Rev. B} {\bf 2002}, {\em 66}, 054407.

\bibitem{kowa20}
Kowalewska, P., Sza\l{}owski, K.
Magnetocaloric properties of V6 molecular magnet.
{\em J. Magn. Magn. Mater.} {\bf 2020}, {\em 496}, 165933.

\bibitem{fuji18}
Fujihala, M. , Sugimoto, T., Tohyama, T., Mitsuda, S., Mole, R. A., Yu, D. H., Yano, S., Inagaki, Y., Morodomi, H., Kawae, T., Sagayama, H., Kumai, R., Murakami, Y., Tomiyasu, K., Matsuo, A., Kindo, K. 
Cluster-Based Haldane State in an Edge-Shared Tetrahedral Spin-Cluster Chain: Fedotovite K$_2$Cu$_3$O(SO$_4$)$_3$.
{\em Phys. Rev. Lett.} {\bf 2018}, {\em 120}, 077201.

\bibitem{stre18}
Stre\v{c}ka, J., Karl'ov\'a, K.
Magnetization curves and low-temperature thermodynamics of two spin-1/2 Heisenberg edge-shared tetrahedra.
{\em AIP Adv.} {\bf 2018}, {\em 8}, 101403.

\bibitem{furr20}
Furrer, A., Podlesnyak, A., Clemente-Juan, J. M., Pomjakushina, E., G\"udel, H. U.
Spin-coupling topology in the copper hexamer compounds.
{\em Phys. Rev. B} {\bf 2020}, {\em 101}, 224417.

\bibitem{evan10}
Evangelisti, M., Brechin, E.K. 
Recipes for Enhanced Molecular Cooling. 
{\em Dalton Trans.} {\bf 2010}, {\em 39}, 4672--4676.

\bibitem{shar13}
Sharples, J. W., Collison, D.
Coordination compounds and the magnetocaloric effect.
{\em Polyhedron} {\bf 2013}, {\em 54}, 91--103.

\bibitem{garl13}
Garlatti, E., Carretta, S., Schnack, J., Amoretti, G., Santini, P. 
Theoretical Design of Molecular Nanomagnets for Magnetic Refrigeration. 
{\em Appl. Phys. Lett.} {\bf 2013}, {\em 103}, 202410. 

\bibitem{liuj14}
Liu, J. L., Chen, Y. C., Guo, F. S., Tong, M. L. 
Recent Advances in the Design of Magnetic Molecules for Use as Cryogenic Magnetic Coolants. 
{\em Coord. Chem. Rev.} {\bf 2014}, {\em 281}, 26--49.

\bibitem{zhen14}
Zheng, Y. Z., Zhou, G. J., Zheng, Z., Winpenny, R. E. P. 
Molecule-Based Magnetic Coolers. 
{\em Chem. Soc. Rev.} {\bf 2014}, {\em 43}, 1462--1475.

\bibitem{evan05}
Evangelisti, M., Candini, A., Ghirri, A., Affronte, M., Brechin, E. K., McInnes, E. J. L. 
Spin-enhanced magnetocaloric effect in molecular nanomagnets, 
{\em Appl. Phys. Lett.} {\bf 2005}, {\em 87}, 072504.

\bibitem{evan09}
Evangelisti, M., Candini, A., Affronte, M., Pasca, E., de Jongh, L. J., Scott, R. T. W., Brechin, E. K.
Magnetocaloric effect in spin-degenerated molecular nanomagnets,
{\em Phys. Rev. B} {\bf 2009}, {\em 79}, 104414.

\bibitem{affr04}        
Affronte, M., Ghirri, A., Carretta, S., Amoretti, G., Piligkos, S., Timco, G.A., Winpenny, R.E.P.
Engineering Molecular Rings for Magnetocaloric Effect. 
{\em Appl. Phys. Lett.} {\bf 2004}, {\em 84}, 3468--3470.

\bibitem{shar14}
Sharples, J. W., Collison, D., McInnes, E. J. L., Schnack, J., Palacios, E., Evangelisti, M. 
Quantum Signatures of a Molecular Nanomagnet in Direct Magnetocaloric Measurements. 
{\em Nat. Commun.} {\bf 2014}, {\em 5}, 1--6.

\bibitem{fuxu14}
Fu, Z., Xiao, Y., Su, Y., Zheng, Y., K\"ogerler, P., Br\"uckel, T. 
Low-lying magnetic excitations and magnetocaloric effect of molecular magnet K$_6$[V$_{15}$As$_6$O$_{42}$(H$_2$O)] $\cdot$ 8H$_2$O.
{\em EPL} {\bf 2015}, {\em 112}, 27003.

\bibitem{pine16}
Pineda, E. M., Lorusso, G., Zangana, K. H., Palacios, E., Schnack, J., Evangelisti, M., Winpenny, R. E. P., McInnes, E. J. L.
Observation of the influence of dipolar and spin frustration effects on the magnetocaloric properties of a trigonal prismatic \{Gd7\} molecular nanomagnet.
{\em Chem. Sci.} {\bf 2016}, {\em 7}, 4891--4895.

\bibitem{fitt19}
Fitta, M., Pe\l{}ka, R., Konieczny, P., Ba\l{}anda, M. 
Multifunctional Molecular Magnets: Magnetocaloric Effect in Octacyanometallates. 
{\em Crystals} {\bf 2019}, {\em 9}, 9.

\bibitem{escu98}
Escuer, A., Vicente, R., Kumara, S. R., Mautner, F. A.
Spin frustration in the butterfly-like tetrameric [Ni$_4$($\mu$-CO$_3$)$_2$(aetpy)$_8$][ClO$_4$]$_4$ (aetpy = (2-aminoethyl)pyridine) complex. Structure and magnetic properties.
{\em J. Chem. Soc., Dalton Trans.} {\bf 1998}, 3473--3477.

\bibitem{hagi06}
Hagiwara, M., Narumi, Y., Matsuo, A., Yashiro, H., Kimura, S., Kindo, K.
Magnetic properties of a Ni tetramer with a butterfly structure in high magnetic fields.
{\em New J. Phys.} {\bf 2006}, {\em 8}, 176. 

\bibitem{kamb50}
Kambe, K.
On the Paramagnetic Susceptibilities of Some Polynuclear Complex Salts.
{\em J. Phys. Soc. Jpn.} {\bf 1950}, {\em 5}, 48--51. 

\bibitem{sinn70}
Sinn, E.
Magnetic Exchange in Polynuclear Metal Complexes.
{\em Coord. Chem. Rev.} {\bf 1970}, {\em 5}, 313--347.
\end{thebibliography}
\end{document}